\documentclass{ws-rv9x6}
\usepackage{ws-rv-thm}
\usepackage[square]{ws-rv-van}
\usepackage{simplewick}
\usepackage{relsize}
\usepackage{alphalph}

\usepackage[utf8]{inputenc}
\usepackage{multirow}
\usepackage{amsmath}
\usepackage{amssymb}
\usepackage{esint}

\newtheorem{exercise}{Exercise}[section]

\newcommand\be{\begin{equation}}
\newcommand\ee{\end{equation}}

\newcommand\D{\Delta}
\newcommand\g{\gamma}
\newcommand\Op{\mathcal{O}}

\newif\ifarxivsubmission

\arxivsubmissiontrue
\ifarxivsubmission
  \usepackage{color}
  \definecolor{darkblue}{rgb}{0.1,0.1,.7}
  \usepackage[colorlinks, linkcolor=darkblue, citecolor=darkblue, urlcolor=darkblue, linktocpage]{hyperref}
  \crop[off]{}
  \makeatletter
  \def\@makechapterhead#1{%                                                                                                      
    {\vbox to 110pt{%        %rvs                                                                                              
    \def\thefootnote{\@fnsymbol\c@footnote}%                                                                                   
    \vspace*{37pt}      %VSPACE FROM TRIM SIZE                                                                                 
        \parindent\z@\raggedright\reset@font
        {\centering{%{\CNfont Chapter~\thechapter\par}%                                                                         
         \vskip 0.25in
    \vbox{
    \CTfont #1\par
    }\par}\par}\nobreak\vfill}}}%  %rvs
  \makeatother
\else
  \usepackage[]{hyperref}
\fi

\makeindex

\begin{document}

\ifarxivsubmission
\chapter{TASI Lectures on  AdS/CFT}
\else
\chapter[Introduction to AdS/CFT]{Introduction to AdS/CFT}
\fi

\author[João Penedones]{João Penedones}

\address{
joao.penedones@epfl.ch\\
\vspace{0.2cm}
Fields and Strings Laboratory, Institute of Physics, EPFL\\
Rte de la Sorge, BSP 728,
CH-1015 Lausanne, Switzerland\\
\vspace{0.2cm}
Centro de Física do Porto, Universidade do Porto,   Portugal\\
\vspace{0.2cm}
Theoretical Physics Department, CERN, Geneva, Switzerland
}

\begin{abstract}
We introduce the AdS/CFT correspondence as a natural extension of QFT in a fixed AdS background.
We start by reviewing some general concepts of CFT, including the embedding space formalism. We then consider QFT in a fixed AdS background and show that one can define boundary operators that enjoy  very similar properties as in a CFT, except for the lack of a stress tensor.
Including a dynamical metric in AdS generates a boundary stress tensor and completes the CFT axioms. 
We also discuss some applications of the bulk geometric intuition to strongly coupled QFT. Finally, we end with a review of the main properties of Mellin amplitudes for CFT correlation functions and their uses in the context of AdS/CFT.
\end{abstract}
\body

\makeatletter
\newalphalph{\alphmult}[mult]{\@alph}{26}
\makeatother
\renewcommand{\thefootnote}{\alphmult{\value{footnote}}}

\ifarxivsubmission
  \thispagestyle{empty}
  \makeatletter
  \renewcommand*\l@chapter[2]{}
  \renewcommand*\l@schapter[2]{}
  \renewcommand*\l@author[2]{}
  \makeatother
  \newpage
  \setcounter{page}{1}
  \setcounter{tocdepth}{2}
  \smalltoc
  \tableofcontents
 
  \newpage
\fi

\section{Introduction}

%This a preliminary and incomplete draft of the lecture notes of the course ``Introduction to AdS/CFT'' at TASI 2015. I hope to complete and improve these notes during and after the lectures, incorporating feedback from the students.

The AdS/CFT correspondence \cite{hep-th/9711200, hep-th/9802109, hep-th/9802150} is a well established general approach to quantum gravity.
However, it is often perceived as a particular construction specific to string theory.
In these lectures I will argue that the AdS/CFT correspondence is the most conservative approach to quantum gravity. 
The quick argument goes as follows:
\begin{itemize}
\item  \emph{System in a box} - we work with Anti-de Sitter (AdS) boundary conditions because AdS is the most symmetric box with a boundary.  This is useful to control large IR effects, even without dynamical gravity.
\item \emph{QFT in the box} - Quantum Field Theory (no gravity) in a fixed AdS background leads to the construction of boundary operators that enjoy an associative and convergent Operator Product Expansion (OPE). The AdS isometries act on the boundary operators like the conformal group in one lower dimension.
\item \emph{Boundary stress-tensor from gravitons} - perturbative metric fluctuations around AdS lead to a boundary stress tensor (weakly coupled to the other boundary operators).
\end{itemize}
Starting from these 3 facts it is entirely natural to define quantum gravity with AdS boundary conditions as Conformal Field Theory (CFT) in one lower dimension. 
%It is the simplest way of UV completing the theory preserving unitarity. 
Of course not all CFTs look like gravity in our universe. That requires the size of the box to be much larger than the Planck length and all higher spin particles to be very heavy (relative to the size of the box). 
As we shall see,  these physical requirements imply that the CFT is strongly coupled and therefore hard to find or construct. The major role of string theory is to provide explicit examples of such CFTs like maximally supersymmetric Yang-Mills  (SYM) theory.

There are many benefits that follow from accepting the AdS/CFT perspective.
Firstly, it makes the holographic nature of gravity manifest. For example, one can immediately match the scaling of the CFT entropy density with the  Bekenstein-Hawking entropy of (large) black holes in AdS. Notice that this is a consequence because it was not used as an argument for AdS/CFT in the previous paragraph. More generally, the AdS/CFT perspective let us translate questions about quantum gravity into mathematically well posed questions about CFT. \footnote{It might not be possible to formulate all quantum gravity questions in CFT language. For example, it is unclear if the experience of an observer falling into a black hole in AdS is a CFT observable \cite{arXiv:1207.3123}.}
Another benefit of the gauge/gravity duality is that it gives us a geometric description of  QFT phenomena, which can be very useful to gain physical
intuition and to create phenomenological models.

This introduction to AdS/CFT will not follow the historical order
of scientific developments.
% The lectures are divided into 5 sections.
Section \ref{sec:CFT} reviews general concepts in CFT. This part is not entirely self contained because this topic is discussed in  detail in the chapter
\emph{Conformal Bootstrap}  by David
Simmons-Duffin \cite{arXiv:1602.07982}. \footnote{See also the lecture notes \cite{arXiv:1601.05000, arXiv:1511.04074}.}
The main purpose of this section is to set up notation,
introduce the embedding space formalism and discuss large $N$ factorization.
Section \ref{sec:AdS} deals with Anti-de Sitter (AdS) spacetime. The first goal
here is to gain intuition about particle dynamics in AdS and 
QFT in a fixed AdS background.
From this point-of-view, we will see
that a gravitational theory with AdS boundary conditions naturally
defines a CFT living on its boundary. In section \ref{sec:AdSCFT},
we discuss the AdS/CFT correspondence in more detail and emphasize
its importance for quantum gravity. We also consider what kind of
CFTs have simple AdS duals and the role of string theory. Furthermore,
we discuss several applications of the gauge/gravity duality as a
tool to geometrize QFT effects.
Finally, in section \ref{sec:Mellin}, we introduce the Mellin representation of CFT correlation functions. We explain the analytic properties of Mellin amplitudes and their particular simplicity in the case of holographic CFTs. 

There are many reviews of AdS/CFT available in the literature. Most
of them are complementary to these lecture notes because they discuss
in greater detail concrete examples of AdS/CFT realized in string
theory. I leave here an incomplete list \cite{hep-th/9905111, arXiv:0909.0518, hep-th/0201253,   arXiv:0709.1523, arXiv:0712.0689, arXiv:0901.0935, arXiv:0903.3246,  arXiv:1010.6134, arXiv:1106.4324}
that can be useful to the readers interested in knowing more about
AdS/CFT. 
The lecture notes \cite{JaredAdSCFT} by Jared Kaplan discuss in greater detail many of the ideas presented here.

\section{Conformal Field Theory \label{sec:CFT}}

This section briefly describes the basic concepts necessary to formulate a non-perturbative definition of CFT.
In the last part, we explain in more detail the embedding space formalism for CFT and 't Hooft's large $N$ expansion, which will be very important in the following sections.

\subsection{Conformal Transformations}

For simplicity, in most formulas, we will consider Euclidean signature.
We start by discussing conformal transformations of $\mathbb{R}^{d}$
in Cartesian coordinates,
\be
ds^{2}=\delta_{\mu\nu}dx^{\mu}dx^{\nu}\ .
\ee
A conformal transformation is a coordinate transformation that preserves
the form of the metric tensor up to a scale factor,
\begin{equation}
\delta_{\mu\nu}\frac{d\tilde{x}^{\mu}}{dx^{\alpha}}\frac{d\tilde{x}^{\nu}}{dx^{\beta}}=\Omega^{2}(x)\delta_{\alpha\beta}\ .\label{eq:conftransf}
\end{equation}
In other words, a conformal transformation is a local dilatation. 
 
\begin{exercise} 
Show that, for $d>2$, the most general infinitesimal conformal
transformation is given by $\tilde{x}^{\mu}=x^{\mu}+\epsilon^{\mu}(x)$
with
\begin{equation}
\epsilon^{\mu}(x)=a^{\mu}+\lambda x^{\mu}+m^{\mu\nu}x_{\nu}+x^{2}b^{\mu}-2x^{\alpha}b_{\alpha}x^{\mu}\ .\label{eq:InfinitesimalConfTransf}
\end{equation}
\end{exercise}

In spacetime dimension $d>2$, conformal transformations form the
group $SO(d+1,1)$. The generators $P_{\mu}$ and $M_{\mu\nu}$ correspond
to translation and rotations and they are present in any relativistic
invariant QFT. In addition, we have the generators of dilatations
$D$ and special conformal transformations $K_{\mu}$. It is convenient
to think of the special conformal transformations as the composition
of an inversion followed by a translation followed by another inversion.
Inversion is the conformal transformation%
\footnote{Inversion is outside the component of the conformal group connected
to the identity. Thus, it is possible to have CFTs that are not invariant
under inversion. In fact, CFTs that break parity also break inversion. %
}
\be
x^{\mu}\to\frac{x^{\mu}}{x^{2}}\ .
\ee
\begin{exercise}
Verify that inversion is a conformal transformation.
\end{exercise}

The form of the generators of the conformal algebra acting on functions
can be obtained from 
\be
\phi\left(x^{\mu}+\epsilon^{\mu}(x)\right)=\left[1+i\, a^{\mu}P_{\mu}-\lambda D+\frac{i}{2}m^{\mu\nu}M_{\mu\nu}+i\, b^{\mu}K_{\mu}\right]\phi\left(x^{\mu}\right)\ ,
\ee
which leads to \footnote{We define the dilatation generator $D$ in a non-standard fashion so that it has real eigenvalues in unitary CFTs.}
\begin{align}
P_{\mu}&=-i\partial_{\mu}\ ,\qquad &D&=-x^{\mu}\partial_{\mu}\ ,\\
 M_{\mu\nu}&=-i\left(x_{\mu}\partial_{\nu}-x_{\nu}\partial_{\mu}\right)\ ,\qquad &K_{\mu}&=2ix_{\mu}x^{\nu}\partial_{\nu}-i\, x^{2}\partial_{\mu}\ .
\end{align}
\begin{exercise}
Show that the generators obey the following commutation
relations
\begin{align}
\left[D,P_{\mu}\right]&=P_{\mu}\ ,\qquad\left[D,K_{\mu}\right]=-K_{\mu}\ ,\qquad\left[K_{\mu},P_{\nu}\right]=2\delta_{\mu\nu}D-2i\, M_{\mu\nu}\ ,\nonumber \\
\left[M_{\mu\nu},P_{\alpha}\right]&=i\left(\delta_{\mu\alpha}P_{\nu}-\delta_{\nu\alpha}P_{\mu}\right)\ ,\qquad\left[M_{\mu\nu},K_{\alpha}\right]=i\left(\delta_{\mu\alpha}K_{\nu}-\delta_{\nu\alpha}K_{\mu}\right)\ ,\nonumber \\
\left[M_{\alpha\beta},M_{\mu\nu}\right]&=i\left(\delta_{\alpha\mu}M_{\beta\nu}+\delta_{\beta\nu}M_{\alpha\mu}-\delta_{\beta\mu}M_{\alpha\nu}-\delta_{\alpha\nu}M_{\beta\mu}\right)\ .\label{eq:conformalalgebra}
\end{align}
\end{exercise}

%\footnote{Here $\eta_{\mu\nu}=\delta_{\mu\nu}$ is just the flat Euclidean metric.
%}
%\begin{align}
%\left[D,P_{\mu}\right]&=P_{\mu}\ ,\qquad\left[D,K_{\mu}\right]=-K_{\mu}\ ,\qquad\left[K_{\mu},P_{\nu}\right]=2\eta_{\mu\nu}D-2i\, M_{\mu\nu}\ ,\nonumber \\
%\left[M_{\mu\nu},P_{\alpha}\right]&=i\left(\eta_{\mu\alpha}P_{\nu}-\eta_{\nu\alpha}P_{\mu}\right)\ ,\qquad\left[M_{\mu\nu},K_{\alpha}\right]=i\left(\eta_{\mu\alpha}K_{\nu}-\eta_{\nu\alpha}K_{\mu}\right)\ ,\nonumber \\
%\left[M_{\alpha\beta},M_{\mu\nu}\right]&=i\left(\eta_{\alpha\mu}M_{\beta\nu}+\eta_{\beta\nu}M_{\alpha\mu}-\eta_{\beta\mu}M_{\alpha\nu}-\eta_{\alpha\nu}M_{\beta\mu}\right)\ .\label{eq:conformalalgebra}
%\end{align} 

\subsection{Local Operators}

Local operators are divided into two types: primary and descendant.
Descendant operators are operators that can be written as (linear
combinations of) derivatives of other local operators. Primary operators
can not be written as derivatives of other local operators. Primary
operators at the origin are annihilated by the generators of special
conformal transformations. Moreover, they are eigenvectors of the
dilatation generator and form irreducible representations of the rotation
group $SO(d)$,
\be
\left[K_{\mu},\mathcal{O}(0)\right]=0\ ,\ \ \ \ 
\left[D,\mathcal{O}(0)\right]=\Delta\,\mathcal{O}(0)\ ,\ \ \ \ 
\left[M_{\mu\nu},\mathcal{O}_{A}(0)\right]=\left[M_{\mu\nu}\right]_{A}^{B}\mathcal{O}_{B}(0)\ .
\nonumber
\ee

Correlation functions of scalar primary operators obey
\begin{equation}
\left\langle \mathcal{O}_{1}(\tilde{x}_{1})\dots\mathcal{O}_{n}(\tilde{x}_{n})\right\rangle =\left|\frac{\partial\tilde{x}}{\partial x}\right|_{x_{1}}^{-\frac{\Delta_{1}}{d}}\dots\left|\frac{\partial\tilde{x}}{\partial x}\right|_{x_{n}}^{-\frac{\Delta_{n}}{d}}\left\langle \mathcal{O}_{1}(x_{1})\dots\mathcal{O}_{n}(x_{n})\right\rangle \label{eq:conftransfcorrelators}
\end{equation}
 for all conformal transformations $x\to\tilde{x}$. As explained
above, it is sufficient to impose Poincaré invariance and this transformation
rule under inversion, % $\tilde{x}^{\mu}=\frac{x^{\mu}}{x^{2}}$,
\be
\left\langle \mathcal{O}_{1}\left(\frac{x_{1}}{x_{1}^{2}}\right)\dots\mathcal{O}_{n}\left(\frac{x_{n}}{x_{n}^{2}}\right)\right\rangle =\left(x_{1}^{2}\right)^{\Delta_{1}}\dots\left(x_{n}^{2}\right)^{\Delta_{n}}\left\langle \mathcal{O}_{1}(x_{1})\dots\mathcal{O}_{n}(x_{n})\right\rangle \ . \nonumber
\ee

This implies that vacuum one-point functions $\left\langle \mathcal{O}(x)\right\rangle $
vanish except for the identity operator (which is the unique operator
with $\Delta=0$). It also fixes the form of the two and three point
functions,
\begin{align}
\left\langle \mathcal{O}_{i}(x)\mathcal{O}_{j}(y)\right\rangle &=\frac{\delta_{ij}}{\left(x-y\right)^{2\Delta_{i}}}\ ,\\
\left\langle \mathcal{O}_{1}(x_{1})\mathcal{O}_{2}(x_{2})\mathcal{O}_{3}(x_{3})\right\rangle &=\frac{C_{123}}{\left|x_{12}\right|^{\Delta_{1}+\Delta_{2}-\Delta_{3}}\left|x_{13}\right|^{\Delta_{1}+\Delta_{3}-\Delta_{2}}\left|x_{23}\right|^{\Delta_{2}+\Delta_{3}-\Delta_{1}}}\ ,\nonumber
\end{align}
where we have normalized the operators to have unit two point function.

The four-point function is not fixed by conformal symmetry because
with four points one can construct two independent conformal invariant
cross-ratios
\be
u=z\bar{z}=\frac{x_{12}^{2}x_{34}^{2}}{x_{13}^{2}x_{24}^{2}}\ ,\qquad v=(1-z)(1-\bar{z})=\frac{x_{14}^{2}x_{23}^{2}}{x_{13}^{2}x_{24}^{2}}\ .
\label{crossratios}
\ee
The general form of the four point function is
\be
\left\langle \mathcal{O}(x_{1})\dots\mathcal{O}(x_{4})\right\rangle =\frac{\mathcal{A}(u,v)}{\left(x_{13}^{2}x_{24}^{2}\right)^{\Delta}}\ .
\ee

\subsection{Ward identities}

To define the stress-energy tensor it is convenient to consider the
theory in a general background metric $g_{\mu\nu}$. Formally, we
can write
\be
\left\langle \mathcal{O}_{1}(x_{1})\dots\mathcal{O}_{n}(x_{n})\right\rangle _{g}=\frac{1}{Z[g]}\int[d\phi]e^{-S[\phi,g]}\mathcal{O}_{1}(x_{1})\dots\mathcal{O}_{n}(x_{n})\ ,
\ee
 where $Z[g]=\int[d\phi]e^{-S[\phi,g]}$ is the partition function
for the background metric $g_{\mu\nu}$. Recalling the classical definition
\be
T^{\mu\nu}(x)=-\frac{2}{\sqrt{g}}\frac{\delta S}{\delta g_{\mu\nu}(x)}\ ,
\ee
it is natural to define the quantum stress-energy tensor operator
via the equation
\be
\frac{Z[g+\delta g]}{Z[g]}=1+\frac{1}{2}\int dx\sqrt{g}\delta g_{\mu\nu}(x)\left\langle T^{\mu\nu}(x)\right\rangle _{g}+O(\delta g^{2})\ ,
\ee
and
\begin{align}
 & \left\langle \mathcal{O}_{1}(x_{1})\dots\mathcal{O}_{n}(x_{n})\right\rangle _{g+\delta g}-\left\langle \mathcal{O}_{1}(x_{1})\dots\mathcal{O}_{n}(x_{n})\right\rangle _{g}\nonumber \\
= & \frac{1}{2}\int dx\sqrt{g}\delta g_{\mu\nu}(x)\left[\left\langle T^{\mu\nu}(x)\mathcal{O}_{1}(x_{1})\dots\mathcal{O}_{n}(x_{n})\right\rangle _{g}\right. \label{eq:Tmunufrommetricvariation}\\
&\qquad \qquad \qquad \qquad \left.
-\left\langle T^{\mu\nu}(x)\right\rangle _{g}\left\langle \mathcal{O}_{1}(x_{1})\dots\mathcal{O}_{n}(x_{n})\right\rangle _{g}\right]+O(\delta g^{2})\ .\nonumber 
\end{align}

Under an infinitesimal coordinate transformation $\tilde{x}^{\mu}=x^{\mu}+\epsilon^{\mu}(x)$,
the metric tensor changes $\tilde{g}_{\mu\nu}=g_{\mu\nu}-\nabla_{\mu}\epsilon_{\nu}-\nabla_{\nu}\epsilon_{\mu}$
but the physics should remain invariant. In particular, the partition
function $Z[g]=Z[\tilde{g}]$ and the correlation functions %
\footnote{If the operators are not scalars (e.g. if they are vector operators)
then one also needs to take into account the rotation of their indices.%
}
\be
\left\langle \mathcal{O}_{1}(\tilde{x}_{1})\dots\mathcal{O}_{n}(\tilde{x}_{n})\right\rangle _{\tilde{g}}=\left\langle \mathcal{O}_{1}(x_{1})\dots\mathcal{O}_{n}(x_{n})\right\rangle _{g}\ ,
\ee
do not change.
This leads to the conservation equation $\left\langle \nabla_{\mu}T^{\mu\nu}(x)\right\rangle _{g}$
and 
\begin{align}
&\sum_{i=1}^{n}\epsilon^{\mu}(x_{i})\frac{\partial}{\partial x_{i}^{\mu}}\left\langle \mathcal{O}_{1}(x_{1})\dots\mathcal{O}_{n}(x_{n})\right\rangle _{g} \label{eq:DiffWardIdentity} \\
&=-\int dx\sqrt{g}\epsilon_{\nu}(x)\left\langle \nabla_{\mu}T^{\mu\nu}(x)\mathcal{O}_{1}(x_{1})\dots\mathcal{O}_{n}(x_{n})\right\rangle _{g}\nonumber
\end{align}
for all $\epsilon^{\mu}(x)$ that decays sufficiently fast at infinity.
Thus $\nabla_{\mu}T^{\mu\nu}=0$ up to contact terms.

Correlation functions of primary operators transform homogeneously
under Weyl transformations of the metric %
\footnote{In general, the partition fungion is not invariant in even dimensions.
This is the Weyl anomaly $Z[\Omega^{2}g]=Z[g]e^{-S_{Weyl}[\Omega,g]}$.%
}
\begin{equation}
\left\langle \mathcal{O}_{1}(x_{1})\dots\mathcal{O}_{n}(x_{n})\right\rangle _{\Omega^{2}g}=
\frac{\left\langle \mathcal{O}_{1}(x_{1})\dots\mathcal{O}_{n}(x_{n})\right\rangle _{g}}
{\left[\Omega(x_{1})\right]^{\Delta_{1}}\dots\left[\Omega(x_{n})\right]^{\Delta_{n}}}\ .\label{eq:Weyltransformation}
\end{equation}

\begin{exercise} Show that this transformation rule under local rescalings
of the metric (together with coordinate invariance) implies (\ref{eq:conftransfcorrelators})
under conformal transformations.
\end{exercise}

Consider now an infinitesimal Weyl transformation $\Omega=1+\omega$,
which corresponds to a metric variation $\delta g_{\mu\nu}=2\omega g_{\mu\nu}$.
From (\ref{eq:Tmunufrommetricvariation}) and (\ref{eq:Weyltransformation})
we conclude that 
\begin{align}
 & \sum_{i=1}^{n}\Delta_{i}\,\omega(x_{i})\left\langle \mathcal{O}_{1}(x_{1})\dots\mathcal{O}_{n}(x_{n})\right\rangle _{g} \nonumber\\
= & -\int dx\sqrt{g}\,\omega(x)g_{\mu\nu}\left[\left\langle T^{\mu\nu}(x)\mathcal{O}_{1}(x_{1})\dots\mathcal{O}_{n}(x_{n})\right\rangle _{g}\right. \label{eq:WeylWardIdentity}\\
&\qquad \qquad \qquad \qquad \left.-\left\langle T^{\mu\nu}(x)\right\rangle _{g}\left\langle \mathcal{O}_{1}(x_{1})\dots\mathcal{O}_{n}(x_{n})\right\rangle _{g}\right]\ .\nonumber 
\end{align}

Consider the following codimension 1 integral over the boundary of
a region $B,$ %
\footnote{In the notation of the \emph{Conformal Bootstrap} chapter \cite{arXiv:1602.07982} this is
the topological operator $Q_{\epsilon}[\partial B]$ inserted in the
correlation function $\left\langle \mathcal{O}_{1}(x_{1})\dots\mathcal{O}_{n}(x_{n})\right\rangle _{g}$. %
} 
\begin{align}
I&=\int_{\partial B}dS_{\mu}\epsilon_{\nu}(x)\left[\left\langle T^{\mu\nu}(x)\mathcal{O}_{1}(x_{1})\dots\mathcal{O}_{n}(x_{n})\right\rangle _{g}\right. \label{eq:WardIdentityFlux}\\
&\qquad \qquad \qquad  \qquad \left.-\left\langle T^{\mu\nu}(x)\right\rangle _{g}\left\langle \mathcal{O}_{1}(x_{1})\dots\mathcal{O}_{n}(x_{n})\right\rangle _{g}\right]\ .\nonumber
\end{align}
One can think of this as the total flux of the current $\epsilon_{\nu}T^{\mu\nu}$,
where $\epsilon_{\nu}(x)$ is an infinitesimal conformal transformation.
Gauss law tells us that this flux should be equal to the integral
of the divergence of the current
\be
\nabla_{\mu}\left(\epsilon_{\nu}T^{\mu\nu}\right)=\epsilon_{\nu}\nabla_{\mu}T^{\mu\nu}+ \nabla_{\mu}\epsilon_{\nu} T^{\mu\nu}=\epsilon_{\nu}\nabla_{\mu}T^{\mu\nu}+\frac{1}{d}\nabla_{\alpha}\epsilon^{\alpha}g_{\mu\nu}T^{\mu\nu}\ ,
\ee
where we used the symmetry of the stress-energy tensor $T^{\mu\nu}=T^{\nu\mu}$
and the definition of an infinitesimal conformal transformation $\nabla_{\mu}\epsilon_{\nu}+\nabla_{\nu}\epsilon_{\mu}=\frac{2}{d}\nabla_{\alpha}\epsilon^{\alpha}g_{\mu\nu}$.
Using Gauss law and (\ref{eq:DiffWardIdentity}) and (\ref{eq:WeylWardIdentity})
we conclude that
\begin{equation}
I=-\sum_{x_{i}\in B}\left[\epsilon^{\mu}(x_{i})\frac{\partial}{\partial x_{i}^{\mu}}+\frac{\Delta_{i}}{d}\nabla_{\alpha}\epsilon^{\alpha}(x_{i})\right]\left\langle \mathcal{O}_{1}(x_{1})\dots\mathcal{O}_{n}(x_{n})\right\rangle _{g}\ .\label{eq:WardIndentityTransf}
\end{equation}
The equality of (\ref{eq:WardIdentityFlux}) and (\ref{eq:WardIndentityTransf})
for any infinitesimal conformal transformation (\ref{eq:InfinitesimalConfTransf})
is the most useful form of the conformal Ward identities.

\begin{exercise} 
Conformal symmetry fixes the three-point function
of a spin 2 primary operator and two scalars up to an overall constant,
\footnote{You can try to derive this formula using the embedding space formalism
of section \ref{sub:Embedding-Space-Formalism}.%
}
\begin{equation}
\langle\mathcal{O}(x_{1})\mathcal{O}(x_{2})T^{\mu\nu}(x_{3})\rangle=C_{12T}\frac{H^{\mu\nu}(x_{1},x_{2},x_{3})}{|x_{12}|^{2\Delta-d+2}|x_{13}|^{d-2}|x_{23}|^{d-2}}\ ,\label{eq:3ptOOT}
\end{equation}
where
\be
H^{\mu\nu}=V^{\mu}V^{\nu}-\frac{1}{d}V_{\alpha}V^{\alpha}\delta^{\mu\nu}\ ,\qquad V^{\mu}=\frac{x_{13}^{\mu}}{x_{13}^{2}}-\frac{x_{23}^{\mu}}{x_{23}^{2}}\ .
\ee
Write the conformal Ward identity (\ref{eq:WardIdentityFlux})=(\ref{eq:WardIndentityTransf})
for the three point function $\langle T^{\mu\nu}(x)\mathcal{O}(0)\mathcal{O}(y)\rangle$
for the case of an infinitesimal dilation $\epsilon^{\mu}(x)=\lambda x^{\mu}$
and with the surface $\partial B$ being a sphere centred at the origin
and with radius smaller than $|y|$. Use this form of the conformal
Ward identity in the limit of an infinitesimally small sphere $\partial B$
and formula (\ref{eq:3ptOOT}) for the three point function to derive
\be
C_{\mathcal{OO}T}=-\frac{d\Delta}{d-1}\frac{1}{S_{d}}\ ,
\ee
where $S_{d}=\frac{2\pi^{d/2}}{\Gamma(d/2)}$ is the volume of a $(d-1)$-dimensional
unit sphere.
\end{exercise}

\subsection{State-Operator Map}

Consider $\mathbb{R}^{d}$ in spherical coordinates. Writing the radial
coordinate as $r=e^{\tau}$ we find
\be
ds^{2}=dr^{2}+r^{2}d\Omega_{d-1}^{2}=e^{2\tau}\left(d\tau^{2}+d\Omega_{d-1}^{2}\right)\ .
\ee
Thus, the cylinder $\mathbb{R}\times S^{d-1}$ can be obtained as
a Weyl transformation of euclidean space $\mathbb{R}^{d}$. 
\begin{exercise} Compute the two-point function of a scalar primary
operator on the cylinder using the Weyl transformation property (\ref{eq:Weyltransformation}).
\end{exercise}

A local operator inserted at the origin of $\mathbb{R}^{d}$ prepares
a state at $\tau=-\infty$ on the cylinder. On the other hand, a state
on a constant time slice of the cylinder can be propagated backwards
in time until it corresponds to a boundary condition on a arbitrarily
small sphere around the origin of $\mathbb{R}^{d}$, which defines
a local operator. Furthermore, time translations on the cylinder correspond
to dilatations on $\mathbb{R}^{d}$. This teaches us that the spectrum
of the dilatation generator on $\mathbb{R}^{d}$ is the same as the
energy spectrum for the theory on $\mathbb{R}\times S^{d-1}$. \footnote{
More precisely, there can be a constant shift equal to the Casimir energy of the vacuum on $S^{d-1}$, which is related with the Weyl anomaly. In $d=2$, this gives the usual energy spectrum $\left(\Delta-\frac{c}{12}\right)\frac{1}{L}$ where $c$ is the central charge and $L$ is the radius of $S^1$.
}

\subsection{Operator Product Expansion}

The Operator Product Expansion (OPE) between two scalar primary operators
takes the following form
\be
\mathcal{O}_{i}(x)\mathcal{O}_{j}(0)=\sum_{k}C_{ijk}|x|^{\Delta_{k}-\Delta_{i}-\Delta_{j}}\left[\mathcal{O}_{k}(0)+\underbrace{\beta\, x^{\mu}\partial_{\mu}\mathcal{O}_{k}(0)+\dots}_{{\rm descendants}}\right]
\ee
where $\beta$ denotes a number determined by conformal symmetry. 
For simplicity we show only the contribution of a scalar operator $\mathcal{O}_{k}$. In general, in the OPE of two scalars there are primary operators of all spins.

\begin{exercise} 
Compute $\beta$ by using this OPE inside a three-point
function.
\end{exercise}

The OPE has a finite radius of convergence inside correlation functions.
This follows from the state operator map with an appropriate choice
of origin for radial quantization.

\subsection{Conformal Bootstrap}

Using the OPE successively one can reduce any $n-$point function
to a sum of one-point functions, which all vanish except for the identity
operator. Thus, knowing the operator content of the theory, \emph{i.e.}
the scaling dimensions $\Delta$ and $SO(d)$ irreps $\mathcal{R}$
of all primary operators, and the OPE coefficients $C_{ijk}$,%
\footnote{For primary operators $\mathcal{O}_{1},\,\mathcal{O}_{2},\,\mathcal{O}_{3}$
transforming in non-trivial irreps of $SO(d)$ there are several OPE
coefficients $C_{123}$. The number of OPE coefficients $C_{123}$
is given by the number of symmetric traceless tensor representations
that appear in the tensor product of the 3 irreps of $SO(d)$ associated
to $\mathcal{O}_{1},\,\mathcal{O}_{2}$ and $\mathcal{O}_{3}$.%
} one can determine all correlation functions of local operators. This
set of data is called CFT data because it essentially defines the
theory. %
\footnote{However, there are observables besides the vacuum correlation functions of
local operators. It is also interesting to study non-local operators
(line operators, surface operators, boundary conditions, etc) and
correlation functions in spaces with non-trivial topology (for example,
correlators at finite temperature). %
} The CFT data is not arbitrary, it must satisfy several constraints:
\begin{itemize}
\item {\bf OPE associativity} - Different ways of using the OPE to compute a correlation
function must give the same result. This leads to the conformal bootstrap
equations described below.
\item {\bf Existence of stress-energy tensor} - The stress-energy tensor $T_{\mu\nu}$
is a conserved primary operator (with $\Delta=d$) whose correlation
functions obey the conformal Ward identities.
\item {\bf Unitarity} - In our Euclidean context this corresponds to reflection
positivity and it implies lower bounds on the scaling dimensions.
It also implies that one can choose a basis of real operators where
all OPE coefficients are real. In the context of statistical physics, there are interesting non-unitary CFTs.
\end{itemize}
It is sufficient to impose OPE associativity for all four-point functions
of the theory. For a four-point function of scalar operators, the
bootstrap equation reads
\be
\sum_{k}C_{12k}C_{k34}G_{\Delta_{k},l_{k}}^{(12)(34)}(x_{1},\dots,x_{4})=\sum_{q}C_{13q}C_{q24}G_{\Delta_{q},l_{q}}^{(13)(24)}(x_{1},\dots,x_{4}) \,, \nonumber
\ee
where $G_{\Delta,l}$ are conformal blocks, which encode the contribution from
 a primary operator of dimension $\Delta$ and spin $l$ and all
its descendants.

\subsection{Embedding Space Formalism\label{sub:Embedding-Space-Formalism}}

The conformal group $SO(d+1,1)$ acts naturally on the space of light
rays through the origin of $\mathbb{R}^{d+1,1}$,
\be
-\left(P^{0}\right)^{2}+\left(P^{1}\right)^{2}+\dots+\left(P^{d+1}\right)^{2}=0\ .
\ee
A section of this light-cone is a $d-$dimensional manifold where
the CFT lives. For example, it is easy to see that the Poincaré section
$P^{0}+P^{d+1}=1$ is just $\mathbb{R}^{d}$. To see this parametrize
this section using
\begin{equation}
P^{0}(x)=\frac{1+x^{2}}{2}\ ,\qquad P^{\mu}(x)=x^{\mu}\ ,\qquad P^{d+1}(x)=\frac{1-x^{2}}{2}\ ,\label{eq:boundAdSPoincareEmb}
\end{equation}
with $\mu=1,\dots,d$ and $x^{\mu}\in\mathbb{R}^{d}$ and compute
the induced metric. In fact, any conformally flat manifold can be
obtained as a section of the light-cone in the embedding space $\mathbb{R}^{d+1,1}$.
Using the parametrization $P^{A}=\Omega(x)P^{A}(x)$ with $x^{\mu}\in\mathbb{R}^{d}$,
one can easily show that the induced metric is simply given by $ds^{2}=\Omega^{2}(x)\delta_{\mu\nu}dx^{\mu}dx^{\nu}$.
With this is mind, it is natural to extend a primary operator from
the physical section to the full light-cone with the following homogeneity
property
\begin{equation}
\mathcal{O}(\lambda P)=\lambda^{-\Delta}\mathcal{O}(P)\ ,\qquad\lambda\in\mathbb{R}\ .\label{eq:EmbHomogeneity}
\end{equation}
This implements the Weyl transformation property (\ref{eq:Weyltransformation}).
One can then compute correlation functions directly in the embedding
space, where the constraints of conformal symmetry are just homogeneity
and $SO(d+1,1)$ Lorentz invariance. Physical correlators are simply
obtained by restricting to the section of the light-cone associated
with the physical space of interest.
This idea goes back to Dirac \cite{Dirac:1936fq} and has been further develop by many authors \cite{Mack:1969rr, Boulware:1970ty, Ferrara:1973eg, Ferrara:1973yt, Cornalba:2009ax, Weinberg:2010fx, Costa:2011mg}.

\begin{exercise} Rederive the form of two and three point functions
of scalar primary operators in $\mathbb{R}^{d}$ using the embedding
space formalism.
\end{exercise} 

Vector primary operators can also be extended to the embedding space.
In this case, we impose
\be
P^{A}\mathcal{O}_{A}(P)=0\ ,\qquad\mathcal{O}_{A}(\lambda P)=\lambda^{-\Delta}\mathcal{O}_{A}(P)\ ,\qquad\lambda\in\mathbb{R}\ ,
\ee
and the physical operator is obtained by projecting the indices to
the section,
\be
\mathcal{O}_{\mu}(x)=\left.\frac{\partial P^{A}}{\partial x^{\mu}}\mathcal{O}_{A}(P)\right|_{P^{A}=P^{A}(x)}\ .
\ee
Notice that this implies a redundancy: $\mathcal{O}_{A}(P)\to\mathcal{O}_{A}(P)+P_{A}\Lambda(P)$
gives rise to the same physical operator $\mathcal{O}(x)$, for any
scalar function $\Lambda(P)$ such that $\Lambda(\lambda P)=\lambda^{-\Delta-1}\Lambda(P)$.
This redundancy together with the constraint $P^{A}\mathcal{O}_{A}(P)=0$
remove 2 degrees of freedom of the $(d+2)$-dimensional vector $\mathcal{O}_{A}$.
\begin{exercise} Show that the two-point function of vector primary
operators is given by 
\be
\left\langle \mathcal{O}^{A}(P_{1})\mathcal{O}^{B}(P_{2})\right\rangle =const\frac{\eta^{AB}\left(P_{1}\cdot P_{2}\right)-P_{2}^{A}P_{1}^{B}}{\left(-2P_{1}\cdot P_{2}\right)^{\Delta+1}}\ ,
\ee
up to redundant terms.
\end{exercise}
\begin{exercise} 
 Consider the parametrization $P^{A}=\left(P^{0},P^{\mu},P^{d+1}\right)=\left(\cosh\tau,\Omega^{\mu},-\sinh\tau\right)$
of the global section $\left(P^{0}\right)^{2}-\left(P^{d+1}\right)^{2}=1$,
where $\Omega^{\mu}$ ($\mu=1,\dots,d$) parametrizes a unit $(d-1)-$dimensional
sphere, $\Omega\cdot\Omega=1$. Show that this section has the geometry
of a cylinder exactly like the one used for the state-operator map. 
\end{exercise}

Conformal correlation functions extended to the light-cone of $\mathbb{R}^{1,d+1}$
are annihilated by the generators of $SO(1,d+1)$
\begin{equation}
\sum_{i=1}^{n}J_{AB}^{(i)}\left\langle \mathcal{O}_{1}(P_{1})\dots\mathcal{O}_{n}(P_{n})\right\rangle =0\ ,\label{eq:EmbeddingSymmetry}
\end{equation}
where $J_{AB}^{(i)}$ is the generator 
\be
J_{AB}=-i\left(P_{A}\frac{\partial}{\partial P^{B}}-P_{B}\frac{\partial}{\partial P^{A}}\right)\ ,
\ee
acting on the point $P_{i}$. For a given choice of light cone section,
some generators will preserve the section and some will not. The first
are Killing vectors (isometry generators) and the second are conformal
Killing vectors. The commutation relations give the usual Lorentz
algebra
\be
\left[J_{AB},J_{CD}\right]=i\left(\eta_{AC}J_{BD}+\eta_{BD}J_{AC}-\eta_{BC}J_{AD}-\eta_{AD}J_{BC}\right)\ .
\label{EmbAlgebra}
\ee
\begin{exercise}
 Check that the conformal algebra (\ref{eq:conformalalgebra})
follows from  (\ref{EmbAlgebra}) and 
\begin{align}
D&=-i\, J_{0,d+1}\ ,\qquad &P_{\mu}&=J_{\mu0}-J_{\mu,d+1}\ , \nonumber\\
M_{\mu\nu}&=J_{\mu\nu}\ , &K_{\mu}&=J_{\mu0}+J_{\mu,d+1}.\label{eq:Embconformalgenerators}
\end{align}
\end{exercise}
\begin{exercise}
Show that equation (\ref{eq:EmbeddingSymmetry})
for $J_{AB}=J_{0,d+1}$ implies time translation invariance on the
cylinder
\be
\sum_{i=1}^{n}\frac{\partial}{\partial\tau_{i}}\left\langle \mathcal{O}_{1}(\tau_{1},\Omega_{1})\dots\mathcal{O}_{n}(\tau_{n},\Omega_{n})\right\rangle =0\ ,
\ee
and dilatation invariance on $\mathbb{R}^{d}$ 
\be
\sum_{i=1}^{n}\left(\Delta_{i}+x_{i}^{\mu}\frac{\partial}{\partial x_{i}^{\mu}}\right)\left\langle \mathcal{O}_{1}(x_{1})\dots\mathcal{O}_{n}(x_{n})\right\rangle =0\ .
\ee
In this case, you will need to use the differential form of the homogeneity property $P^{A}\frac{\partial}{\partial P^{A}}\mathcal{O}_{i}(P)=-\Delta_{i}\mathcal{O}_{i}(P)$.
It is instructive to do this exercise for the other generators as
well.
\end{exercise}

\subsection{Large N Factorization}

\begin{figure}
\begin{centering}
\includegraphics[clip,width=0.9\textwidth]{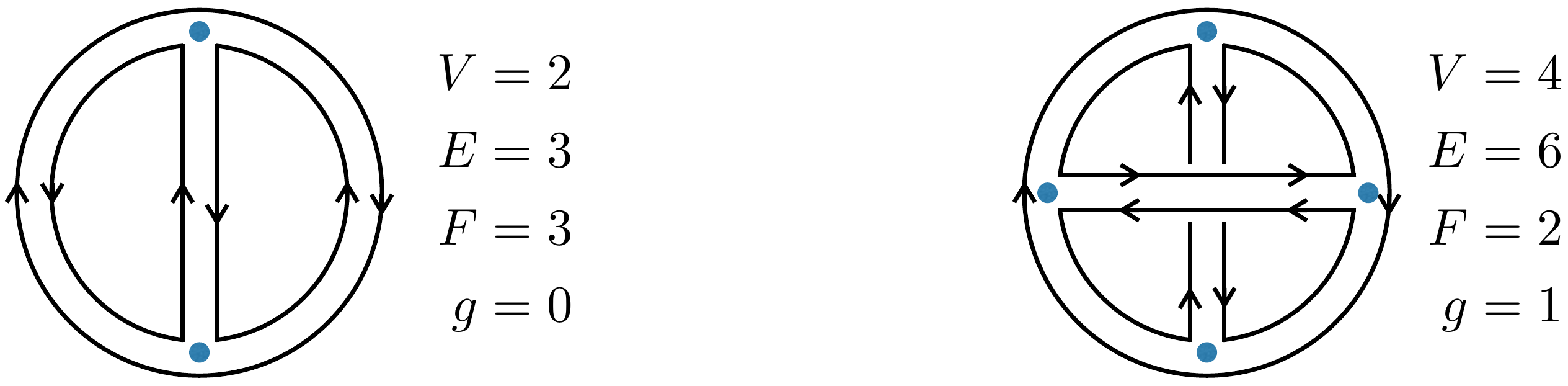}
\caption{Vacuum diagrams in the double line notation. Interaction vertices are marked with a small blue dot. The left diagram is planar while the diagram on the right has the topology of a torus (genus 1 surface).
\label{fig:doublelinediagrams}}
\end{centering}
\end{figure}

Consider a $U(N)$ gauge theory with fields valued in the adjoint
representation. Schematically, we can write the action as
\be
S=\frac{N}{\lambda}\int dx\ {\rm Tr}\left[\left(D\Phi\right)^{2}+c_{3}\Phi^{3}+c_{4}\Phi^{4}+\dots\right]
\ee
where we introduced the 't Hooft coupling $\lambda=g_{YM}^{2}N$ and
$c_{i}$ are other coupling constants independent of $N$. Following 't Hooft \cite{'tHooft:1973jz},
we consider the limit of large $N$ with $\lambda$ kept fixed. The
propagator of an adjoint field obeys
\be
\left\langle \Phi_{j}^{i}\Phi_{l}^{k}\right\rangle \propto\frac{\lambda}{N}\delta_{l}^{i}\delta_{j}^{k}
\ee
where we used the fact that the adjoint representation can be represented
as the direct product of the fundamental and the anti-fundamental
representation. This suggests that one can represent a propagator
by a double line, where each line denotes the flow of a fundamental
index. Start by considering the vacuum diagrams in this language.
A diagram with $V$ vertices, $E$ propagators (or edges) and $F$
lines (or faces) scales as
\be
\left(\frac{N}{\lambda}\right)^{V}\left(\frac{\lambda}{N}\right)^{E}N^{F}=\left(\frac{N}{\lambda}\right)^{\chi}\lambda^{F}\ ,
\ee
where $\chi=V+F-E=2-2g$ is the minimal Euler character of the two dimensional
surface where the double line diagram can be embedded and $g$ is
the number of handles of this surface. Therefore, the large $N$ limit
is dominated by diagrams that can be drawn on a sphere $(g=0)$. These
diagrams are called planar diagrams. For a given topology, there is
an infinite number of diagrams that contribute with increasing powers
of the coupling $\lambda$, corresponding to tesselating the surface
with more and more faces. Figure \ref{fig:doublelinediagrams} shows two examples of vacuum diagrams in the double line notation.
This topological expansion has the structure
of string perturbation theory with $\lambda/N$ playing the role of
the string coupling. As we shall see this is precisely realized in
maximally supersymmetric Yang-Mills theory (SYM).

Let us now consider single-trace local operators of the form $\mathcal{O}=c_{J}{\rm Tr}\left(\Phi^{J}\right)$,
where $c_{J}$ is a normalization constant independent of $N$. Adapting
the argument above, it is easy to conclude that the connected correlators
are given by a large $N$ expansion of the form
\be
\left\langle \mathcal{O}_{1}\dots\mathcal{O}_{n}\right\rangle _{c}=\sum_{g=0}^{\infty}N^{2-n-2g}f_{g}(\lambda)\ ,
\ee
which is dominated by the planars diagrams $(g=0)$. Moreover, we
see that the planar two-point function is independent of $N$ while
connected higher point functions are suppressed by powers of $N$.
This is large $N$ factorization. In particular it implies that the
two-point function of a multi-trace operator $\tilde{\mathcal{O}}(x)=:\mathcal{O}_{1}(x)\dots\mathcal{O}_{k}(x):$
is dominated by the product of the two-point functions of its single-trace
constituents
\be
\left\langle \tilde{\mathcal{O}}(x)\tilde{\mathcal{O}}(y)\right\rangle \approx\prod_{i}\left\langle \mathcal{O}_{i}(x)\mathcal{O}_{i}(y)\right\rangle =\frac{1}{(x-y)^{2\sum_{i}\Delta_{i}}}\ ,
\ee
where we assumed that the single-trace operators were scalar conformal
primaries properly normalized. We conclude that the scaling dimension
of the multi-trace operator $\tilde{\mathcal{O}}$ is given by $\sum_{i}\Delta_{i}+O(1/N^{2})$
. In other words, the space of local operators in a large $N$ CFT
has the structure of a Fock space with single-trace operators playing
the role of single particle states of a weakly coupled theory.
This is the form of large $N$ factorization relevant for AdS/CFT. However, notice that conformal invariance was not important for the argument.
It is well known that large $N$ factorization also occurs in confining gauge theories. Physically, it means that colour singlets (like glueballs or mesons) interact weakly in large $N$ gauge theories (see \cite{Witten:1979kh} for a clear summary).

The stress tensor has a natural normalization that follows from the
action, $T_{\mu\nu}\sim\frac{N}{\lambda}{\rm Tr}\left(\partial_{\mu}\Phi\partial_{\nu}\Phi\right)$.
This leads to the large $N$ scaling
\begin{equation}
\left\langle T_{\mu_{1}\nu_{1}}(x_{1})\dots T_{\mu_{n}\nu_{n}}(x_{n})\right\rangle _{c}\sim N^{2}\ ,\label{eq:TTTTTlargeNscaling}
\end{equation}
which will be important below.
This normalization of $T_{\mu\nu}$ is also fixed by the Ward identities.

\section{Anti-de Sitter Spacetime \label{sec:AdS}}

Euclidean AdS spacetime is the hyperboloid 
\begin{equation}
-\left(X^{0}\right)^{2}+\left(X^{1}\right)^{2}+\dots+\left(X^{d+1}\right)^{2}=-R^{2}\ ,\qquad X^{0}>0\ ,\label{eq:EAdS}
\end{equation}
embedded in $\mathbb{R}^{d+1,1}$. For large values of $X^{0}$ this
hyperboloid approaches the light-cone of the embedding space that
we discussed in section \ref{sub:Embedding-Space-Formalism}. It is
clear from the definition that Euclidean AdS is invariant under $SO(d+1,1)$.
The generators are given by
\be
J_{AB}=-i\left(X_{A}\frac{\partial}{\partial X^{B}}-X_{B}\frac{\partial}{\partial X^{A}}\right)\ .
\ee
Poincaré coordinates are defined by 
\begin{eqnarray}
X^{0} & = & R \frac{1+x^{2}+z^{2}}{2z}\nonumber \\
X^{\mu} & = & R\frac{x^{\mu}}{z}\label{eq:AdSPoincareEmb}\\
X^{d+1} & = & R \frac{1-x^{2}-z^{2}}{2z}\nonumber 
\end{eqnarray}
where $x^{\mu}\in\mathbb{R}^{d}$ and $z>0$. In these coordinates,
the metric reads
\be
ds^{2}=R^{2}\frac{dz^{2}+\delta_{\mu\nu}dx^{\mu}dx^{\nu}}{z^{2}}\ .
\ee
This shows that AdS is conformal to $\mathbb{R}^{+}\times\mathbb{R}^{d}$
whose boundary at $z=0$ is just $\mathbb{R}^{d}$. These coordinates
make explicit the subgroup $SO(1,1)\times ISO(d)$ of the full isometry
group of AdS. These correspond to dilatation and Poincaré symmetries
inside the $d-$dimensional conformal group. In particular, the dilatation
generator is 
\be
D=-i\, J_{0,d+1}=-X_{0}\frac{\partial}{\partial X^{d+1}}+X_{d+1}\frac{\partial}{\partial X^{0}}=-z\frac{\partial}{\partial z}-x^{\mu}\frac{\partial}{\partial x^{\mu}}\ .
\ee

Another useful coordinate system is 
\begin{eqnarray}
X^{0} & = & R\cosh\tau\cosh\rho\nonumber \\
X^{\mu} & = & R\,\Omega^{\mu}\sinh\rho\label{eq:AdSEmbglobalcoord}\\
X^{d+1} & = & -R\sinh\tau\cosh\rho\nonumber 
\end{eqnarray}
where $\Omega^{\mu}$ ($\mu=1,\dots,d$) parametrizes a unit $(d-1)-$dimensional
sphere, $\Omega\cdot\Omega=1$. The metric is given by
\be
ds^{2}=R^{2}\left[\cosh^{2}\rho\, d\tau^{2}+d\rho^{2}+\sinh^{2}\rho\, d\Omega_{d-1}^{2}\right]\ .
\ee
To understand the global structure of this spacetime it is convenient
to change the radial coordinate via $\tanh\rho=\sin r$ so that $r\in[0,\frac{\pi}{2}[$.
Then, the metric becomes
\begin{equation}
ds^{2}=\frac{R^{2}}{\cos^{2}r}\left[d\tau^{2}+dr^{2}+\sin^{2}r\, d\Omega_{d-1}^{2}\right]\ ,\label{eq:EAdSglobalcompactcoord}
\end{equation}
which is conformal to a solid cylinder whose boundary at $r=\frac{\pi}{2}$
is $\mathbb{R}\times S^{d-1}$. In these coordinates, the dilatation
generator $D=-i\, J_{0,d+1}=-\frac{\partial}{\partial\tau}$ is the
hamiltonian conjugate to global time.

\subsection{Particle dynamics in AdS}

For most purposes it is more convenient to work in Euclidean signature
and analytically continue to Lorentzian signature only at the end
of the calculation. However, it is important to discuss the Lorentzian
signature to gain some intuition about real time dynamics. In this
case, AdS is defined by the universal cover of the manifold
\begin{equation}
-\left(X^{0}\right)^{2}+\left(X^{1}\right)^{2}+\dots+\left(X^{d}\right)^{2}-\left(X^{d+1}\right)^{2}=-R^{2}\ ,\label{eq:EmbeddingLorentzianAdS}
\end{equation}
embedded in $\mathbb{R}^{d,2}$. The universal cover means that we
should unroll the non-contractible (timelike) cycle. To see this explicitly
it is convenient to introduce global coordinates%
\footnote{Notice that this is just the analytic continuation $\tau\to i\, t$
and $X^{d+1}\to i\, X^{d+1}$ of the Euclidean global coordinates
(\ref{eq:AdSEmbglobalcoord}).%
}
\begin{eqnarray}
X^{0} & = & R\cos t\cosh\rho \nonumber\\
X^{\mu} & = & R\,\Omega^{\mu}\sinh\rho\\
X^{d+1} & = & -R\sin t\cosh\rho \nonumber
\end{eqnarray}
where $\Omega^{\mu}$ ($\mu=1,\dots,d$) parametrizes a unit $(d-1)-$dimensional
sphere. The original hyperboloid is covered with $t\in[0,2\pi[$ but
we consider $t\in\mathbb{R}$. The metric is given by
\begin{equation}
ds^{2}=R^{2}\left[-\cosh^{2}\rho\, dt^{2}+d\rho^{2}+\sinh^{2}\rho\, d\Omega_{d-1}^{2}\right]\ .\label{eq:globalcoordLorentzian}
\end{equation}

To understand the global structure of this spacetime it is convenient
to change the radial coordinate via $\tanh\rho=\sin r$ so that $r\in[0,\frac{\pi}{2}[$.
Then, the metric becomes
\begin{equation}
ds^{2}=\frac{R^{2}}{\cos^{2}r}\left[-dt^{2}+dr^{2}+\sin^{2}r\, d\Omega_{d-1}^{2}\right]\ ,\label{eq:compactglobalcoords}
\end{equation}
which is conformal to a solid cylinder whose boundary at $r=\frac{\pi}{2}$
is $\mathbb{R}\times S^{d-1}$.

Geodesics are given by the intersection of AdS with 2-planes through
the origin of the embedding space. In global coordinates, the simplest
timelike geodesic describes a particle sitting at $\rho=0$. This
corresponds to (the universal cover of) the intersection of $X^{\mu}=0$
for $\mu=1,\dots,d$ with the hyperboloid (\ref{eq:EmbeddingLorentzianAdS}).
Performing a boost in the $X^{1},X^{d+1}$ plane we can obtain an
equivalent timelike geodesic $X^{1}\cosh\beta=X^{d+1}\sinh\beta$
and $X^{\mu}=0$ for $\mu=2,\dots,d$. In global coordinates, this
gives an oscillating trajectory 
\be
\tanh\rho=\tanh\beta\,\sin t\ ,
\ee
with period $2\pi$. In fact, all timelike geodesics oscillate with
period $2\pi$ in global time. One can say AdS acts like a box that
confines massive particles. However, it is a very symmetric box that
does not have a center because all points are equivalent.

Null geodesics in AdS are also null geodesics in the embedding space.
For example, the null ray $X^{d+1}-X^{1}=X^{0}-R=X^{\mu}=0$ for $\mu=2,\dots,d$
is a null ray in AdS which in global coordinates is given by $\cosh\rho=\frac{1}{\cos t}$.
This describes a light ray that passes through the origin at $t=0$
and reaches the conformal boundary $\rho=\infty$ at $t=\pm\frac{\pi}{2}$.
All light rays in AdS start and end at the conformal boundary traveling
for a global time interval equal to $\pi$. 

One can also introduce Poincaré coordinates 
\begin{eqnarray}
X^{\mu} & = & R\frac{x^{\mu}}{z} \nonumber \\
X^{d} & = & \frac{R}{2}\frac{1-x^{2}-z^{2}}{z}\\
X^{d+1} & = & \frac{R}{2}\frac{1+x^{2}+z^{2}}{z} \nonumber
\end{eqnarray}
where now $\mu=0,1,\dots,d-1$ and $x^{2}=\eta_{\mu\nu}x^{\mu}x^{\nu}$.
However, in Lorentzian signature, Poincaré coordinates do not cover
the entire spacetime. Surfaces of constant $z$ approach the light-like
surface $X^{d}+X^{d+1}=0$ when $z\to\infty$. This null surface is often called the Poincaré horizon.

We have seen that AdS acts like a box for classical massive particles.
Quantum mechanically, this confining potential gives rise to a discrete
energy spectrum. Consider the Klein-Gordon equation
\be
\nabla^{2}\phi=m^{2}\phi\ ,
\ee
in global coordinates (\ref{eq:globalcoordLorentzian}). In order
to emphasize the correspondence with CFT we will solve this problem
using an indirect route. Firstly, consider the action of the quadratic
Casimir of the AdS isometry group on a scalar field
\begin{equation}
\frac{1}{2}J_{AB}J^{BA}\phi=\left[-X^{2}\partial_{X}^{2}+X\cdot\partial_{X}\left(d+X\cdot\partial_{X}\right)\right]\phi\ .\label{eq:CasimirEmbedding}
\end{equation}
Formally, we are extending the function $\phi$ from AdS, defined
by the hypersurface $X^{2}=-R^{2}$, to the embedding space. However,
the action of the quadratic Casimir is independent of this extension
because the generators $J_{AB}$ are interior to AdS, i.e. $\left[J_{AB},X^{2}+R^{2}\right]=0$.
If we foliate the embedding space $\mathbb{R}^{d,2}$ with AdS surfaces
of different radii $R$, we obtain that the laplacian in the embedding
space can be written as
\be
\partial_{X}^{2}=-\frac{1}{R^{d+1}}\frac{\partial}{\partial R}R^{d+1}\frac{\partial}{\partial R}+\nabla_{AdS}^{2}\ .
\ee
Substituting this in (\ref{eq:CasimirEmbedding}) and noticing that
$X\cdot\partial_{X}=R\partial_{R}$ we conclude that
\begin{equation}
\frac{1}{2}J_{AB}J^{BA}\phi=R^{2}\nabla_{AdS}^{2}\phi\ .\label{eq:CasimirAdS}
\end{equation}
Therefore, we should identify $m^{2}R^{2}$ with the quadratic Casimir
of the conformal group.

The Lorentzian version of the conformal generators (\ref{eq:Embconformalgenerators})
is 
\begin{align}
D&=-J_{0,d+1}\ ,\qquad &P_{\mu}&=J_{\mu0}+i\, J_{\mu,d+1}\ ,\\
 M_{\mu\nu}&=J_{\mu\nu} \ , \qquad &K_{\mu}&=J_{\mu0}-i\, J_{\mu,d+1}\ .
\end{align}
\begin{exercise} 
Show that, in global coordinates, the conformal generators
take the form
\begin{align*}
D&=i\frac{\partial}{\partial t}\ ,\qquad M_{\mu\nu}=-i\left(\Omega_{\mu}\frac{\partial}{\partial\Omega^{\nu}}-\Omega_{\nu}\frac{\partial}{\partial\Omega^{\mu}}\right)\ ,\\
P_{\mu}&=-ie^{-it}\left[\Omega_{\mu}\left(\partial_{\rho}-i\tanh\rho\,\partial_{t}\right)+\frac{1}{\tanh\rho}\nabla_{\mu}\right]\ ,\\
K_{\mu}&=ie^{it}\left[\Omega_{\mu}\left(-\partial_{\rho}-i\tanh\rho\,\partial_{t}\right)-\frac{1}{\tanh\rho}\nabla_{\mu}\right]\ ,
\end{align*}
where $\nabla_{\mu}=\frac{\partial}{\partial\Omega^{\mu}}-\Omega_{\mu}\Omega^{\nu}\frac{\partial}{\partial\Omega^{\nu}}$
is the covariant derivative on the unit sphere $S^{d-1}$. 
\end{exercise}

In analogy with the CFT construction we can look for primary states,
which are annihilated by $K_{\mu}$ and are eigenstates of the hamiltonian,
$D\phi=\Delta\phi$. The condition $K_{\mu}\phi=0$ splits in one
term proportional to $\Omega_{\mu}$ and one term orthogonal to $\Omega_{\mu}$.
The second term implies that $\phi$ is independent of the angular
variables $\Omega^{\mu}$. The first term gives $\left(\partial_{\rho}+\Delta\tanh\rho\right)\phi=0$,
which implies that 
\begin{equation}
\phi\propto\left(\frac{e^{-it}}{\cosh\rho}\right)^{\Delta}=\left(\frac{R}{X^{0}-X^{d+1}}\right)^{\Delta}\ .\label{eq:groundstateWFinAdS}
\end{equation}
This is the lowest energy state. One can get excited states acting
with $P_{\mu}$. Notice that all this states will have the same value
of the quadratic Casimir 
\be
\frac{1}{2}J_{AB}J^{BA}\phi=\Delta(\Delta-d)\phi\ .
\ee
This way one can generate all normalizable solutions of $\nabla^{2}\phi=m^{2}\phi$
with $m^{2}R^{2}=\Delta(\Delta-d)$. This shows that the one-particle
energy spectrum is given by $\omega=\Delta+l+2n$ where $l=0,1,2,\dots$
is the spin, generated by acting with $P_{\mu_{1}}\dots P_{\mu_{l}}-traces\ $,
and $n=0,1,2,\dots$ is generated by acting with $\left(P^{2}\right)^{n}$.

\begin{exercise}
Given the symmetry of the metric (\ref{eq:EAdSglobalcompactcoord})
we can look for solutions of the form 
\be
\phi=e^{i\omega t}Y_{l}(\Omega)F(r)\ ,
\ee
where $Y_{l}(\Omega)$ is a spherical harmonic with eigenvalue $-l(l+d-2)$
of the laplacian on the unit sphere $S^{d-1}$. Derive a differential
equation for $F(r)$ and show that it is solved by 
\be
F(r)=\left(\cos r\right)^{\Delta}\left(\sin r\right)^{l}\ _{2}F_{1}\left(\frac{l+\Delta-\omega}{2},\frac{l+\Delta+\omega}{2},l+\frac{d}{2},\sin r\right)\ ,
\ee
with $2\Delta=d+\sqrt{d^{2}+4(mR)^{2}}$. We chose this solution because
it is smooth at $r=0$. Now we also need to impose another boundary
condition at the boundary of AdS $r=\frac{\pi}{2}$. Imposing that
there is no energy flux through the boundary leads to the quantization
of the energies $|\omega|=\Delta+l+2n$ with $n=0,1,2,\dots$ (see reference 
\cite{hep-th/9905111}).
\end{exercise}

If there are no interactions between the particles in AdS, then the
Hilbert space is a Fock space and the energy of a multi-particle state
is just the sum of the energies of each particle. Turning on small
interactions leads to small energy shifts of the multi-particle states.
This structure is very similar to the space of local operators in
large $N$ CFTs if we identify single-particle states with single-trace
operators.

\subsection{Quantum Field Theory in AdS}

Let us now return to Euclidean signature and consider QFT on the AdS
background. For simplicity, consider a free scalar field with action
\be
S=\int_{AdS}dX\left[\frac{1}{2}\left(\nabla\phi\right)^{2}+\frac{1}{2}m^{2}\phi^{2}\right]\ .
\ee
The two-point function $\left\langle \phi(X)\phi(Y)\right\rangle $
is given by the propagator $\Pi(X,Y)$, which obeys
\be
\left[\nabla_{X}^{2}-m^{2}\right]\Pi(X,Y)=-\delta(X,Y)\ .
\ee
From the symmetry of the problem it is clear that the propagator can
only depend on the invariant $X\cdot Y$ or equivalently on the chordal distance
$\zeta=(X-Y)^{2}/R^{2}$. From now on we will set $R=1$ and all
lengths will be expressed in units of the AdS radius.
\begin{exercise} 
Use (\ref{eq:CasimirEmbedding}) and (\ref{eq:CasimirAdS})
to show that
\be
\Pi(X,Y)=\frac{\mathcal{C}_{\Delta}}{\zeta^{\Delta}}\ _{2}F_{1}\left(\Delta,\Delta-\frac{d}{2}+\frac{1}{2},2\Delta-d+1,-\frac{4}{\zeta}\right)\ ,
\ee
where $2\Delta=d+\sqrt{d^{2}+(2m)^{2}}$ and 
\be
\mathcal{C}_{\Delta}=\frac{\Gamma(\Delta)}{2\pi^{\frac{d}{2}}\Gamma\left(\Delta-\frac{d}{2}+1\right)}\ .
\ee
\end{exercise}

For a free field, higher point functions are simply given by Wick
contractions. For example,
\begin{align}
\left\langle \phi(X_{1})\phi(X_{2})\phi(X_{3})\phi(X_{4})\right\rangle &=\Pi(X_{1},X_{2})\Pi(X_{3},X_{4})
+\Pi(X_{1},X_{3})\Pi(X_{2},X_{4})\nonumber \\
&+\Pi(X_{1},X_{4})\Pi(X_{2},X_{3})\ .\label{eq:Disconnected4ptAdS}
\end{align}
Weak interactions of $\phi$ can be treated perturbatively. Suppose
the action includes a cubic term, 
\begin{equation}
S=\int_{AdS}dX\left[\frac{1}{2}\left(\nabla\phi\right)^{2}+\frac{1}{2}m^{2}\phi^{2}+\frac{1}{3!}g\phi^{3}\right]\ .\label{eq:actionAdSscalar}
\end{equation}
Then, there is a non-vanishing three-point function
\be
\left\langle \phi(X_{1})\phi(X_{2})\phi(X_{3})\right\rangle =-g\int_{AdS}dY\,\Pi(X_{1},Y)\Pi(X_{2},Y)\Pi(X_{3},Y)+O(g^{3})\ , \nonumber
\ee
and a connected part of the four-point function of order $g^{2}$.
This is very similar to perturbative QFT in flat space. 

Given a correlation function in AdS we can consider the limit where
we send all points to infinity. More precisely, we introduce 
\begin{equation}
\mathcal{O}(P)=\frac{1}{\sqrt{C_{\Delta}}}\lim_{\lambda\to\infty}\lambda^{\Delta}\,\phi\left(X=\lambda P+\dots\right),\label{eq:OaslimitofPhi}
\end{equation}
where $P$ is a future directed null vector in $\mathbb{R}^{d+1,1}$
and the $\dots$ denote terms that do not grow with $\lambda$ whose
only purpose is to enforce the AdS condition $X^{2}=-1$. In other
words, the operator $\mathcal{O}(P)$ is the limit of the field $\phi(X)$
when $X$ approaches the boundary point $P$ of AdS. Notice that,
by definition, the operator $\mathcal{O}(P)$ obeys the homogeneity
condition (\ref{eq:EmbHomogeneity}). Correlation functions of $\mathcal{O}$
are naturally defined by the limit of correlation functions of $\phi$
in AdS. For example, the two-point function is given by
\begin{equation}
\left\langle \mathcal{O}(P_{1})\mathcal{O}(P_{2})\right\rangle =\frac{1}{\left(-2P_{1}\cdot P_{2}\right)^{\Delta}}+O(g^{2})\ ,\label{eq:CFT2ptfromAdS}
\end{equation}
which is exactly the CFT two-point function of a primary operator
of dimension $\Delta$. The three-point function $\left\langle \mathcal{O}(P_{1})\mathcal{O}(P_{2})\mathcal{O}(P_{3})\right\rangle$ is given by
\begin{align}
&\label{eq:CFT3ptFromAdS} -g\, C_{\Delta}^{-\frac{3}{2}}\int_{AdS}dX\,\Pi(X,P_{1})\Pi(X,P_{2})\Pi(X,P_{3})
+O(g^{3})
\ , 
\end{align}
where
\be
\Pi(X,P)=\lim_{\lambda\to\infty}\lambda^{\Delta}\,\Pi\left(X,Y=\lambda P+\dots\right)=\frac{C_{\Delta}}{\left(-2P\cdot X\right)^{\Delta}}
\ee
is the bulk to boundary propagator.
\begin{exercise} 
Write the bulk to boundary propagator in Poincaré coordinates.
\end{exercise} 
\begin{exercise} 
\label{ex:3pt}
Compute the following generalization of the integral
in (\ref{eq:CFT3ptFromAdS}), 
\be
\int_{AdS}dX\prod_{i=1}^{3}\frac{1}{\left(-2P_{i}\cdot X\right)^{\Delta_{i}}}\ ,
\ee
and show that it reproduces the expected formula for the CFT three-point
function $\left\langle \mathcal{O}_{1}(P_{1})\mathcal{O}_{2}(P_{2})\mathcal{O}_{3}(P_{3})\right\rangle $.
It is helpful to use the integral representation
\be
\frac{1}{\left(-2P\cdot X\right)^{\Delta}}=\frac{1}{\Gamma(\Delta)}\int_{0}^{\infty}\frac{ds}{s}s^{\Delta}e^{2sP\cdot X}
\ee
to bring the AdS integral to the form
\be
\int_{AdS}dXe^{2X\cdot Q}
\ee
with $Q$ a future directed timelike vector. Choosing the $X^{0}$
direction along $Q$ and using the Poincaré coordinates (\ref{eq:AdSPoincareEmb})
it is easy to show that
\be
\int_{AdS}dXe^{2X\cdot Q}=\pi^{\frac{d}{2}}\int_{0}^{\infty}\frac{dz}{z}\, z^{-\frac{d}{2}}e^{-z+Q^{2}/z}\ .
\ee
To factorize the remaining integrals over $s_{1},s_{2},s_{3}$ and
$z$ it is helpful to change to the variables $t_{1},t_{2},t_{3}$
and $z$ using 
\be
s_{i}=\frac{\sqrt{z\, t_{1}t_{2}t_{3}}}{t_{i}}\ .
\ee
\end{exercise}

\subsubsection{State-Operator Map}

We have seen that the correlation functions of the boundary operator
(\ref{eq:OaslimitofPhi}) have the correct homogeneity property and
$SO(d+1,1)$ invariance expected of CFT correlators of a primary scalar
operator with scaling dimension $\Delta$. We will now argue that
they also obey an associative OPE. The argument is very similar to
the one used in CFT. We think of the correlation functions as
vacuum expectation values of time ordered products
\be
\left\langle \phi(X_{1})\phi(X_{2})\phi(X_{3})\dots\right\rangle =\left\langle 0\right|\dots\hat{\phi}(\tau_{3},\rho_{3},\Omega_{3})\hat{\phi}(\tau_{2},\rho_{2},\Omega_{2})\hat{\phi}(\tau_{1},\rho_{1},\Omega_{1})\left|0\right\rangle \ , \nonumber
\ee
where we assumed $\tau_{1}<\tau_{2}<0<\tau_{3}<\dots$ . We then insert
a complete basis of states at $\tau=0$,
\begin{align}
&\left\langle \phi(X_{1})\phi(X_{2})\phi(X_{3})\dots\right\rangle \\
=&\sum_{\psi}\left\langle 0\right|\dots\hat{\phi}(\tau_{3},\rho_{3},\Omega_{3})\left|\psi\right\rangle \left\langle \psi\right|\hat{\phi}(\tau_{2},\rho_{2},\Omega_{2})\hat{\phi}(\tau_{1},\rho_{1},\Omega_{1})\left|0\right\rangle \ . \nonumber
\end{align}
Using $\hat{\phi}(\tau,\rho,\Omega)=e^{\tau D}\hat{\phi}(0,\rho,\Omega)e^{-\tau D}$
and choosing an eigenbasis of the Hamiltonian $D=-\frac{\partial}{\partial\tau}$
it is clear that the sum converges for the assumed time ordering.
The next step, is to establish a one-to-one map between the states
$\left|\psi\right\rangle $ and boundary operators. It is clear that
every boundary operator (\ref{eq:OaslimitofPhi}) defines a state.
Inserting the boundary operator at $P^{A}=\left(P^{0},P^{\mu},P^{d+1}\right)=\left(\frac{1}{2},0,\frac{1}{2}\right)$,
which is the boundary point defined by $\tau\to-\infty$ in global
coordinates, we can write 
\be
\left\langle \dots\phi(X_{3})\mathcal{O}(P)\right\rangle =\left\langle 0\right|\dots\hat{\phi}(\tau_{3},\rho_{3},\Omega_{3})\left|\mathcal{O}\right\rangle \ ,
\ee
where
\begin{align}
\left|\mathcal{O}\right\rangle &=\lim_{\tau\to-\infty}\left(e^{-\tau}\cosh\rho\right)^{\Delta}\hat{\phi}(\tau,\rho,\Omega)\left|0\right\rangle \\
&=\sum_{\psi}\left|\psi\right\rangle \left(\cosh\rho\right)^{\Delta}\lim_{\tau\to-\infty}\left\langle \psi\right|e^{\tau(D-\Delta)}\hat{\phi}(0,\rho,\Omega)\left|0\right\rangle \ .\nonumber
\end{align}
The limit $\tau\to-\infty$ projects onto the primary state with wave
function (\ref{eq:groundstateWFinAdS}).

\begin{figure}
\begin{centering}
\includegraphics[clip,width=0.6\textwidth]{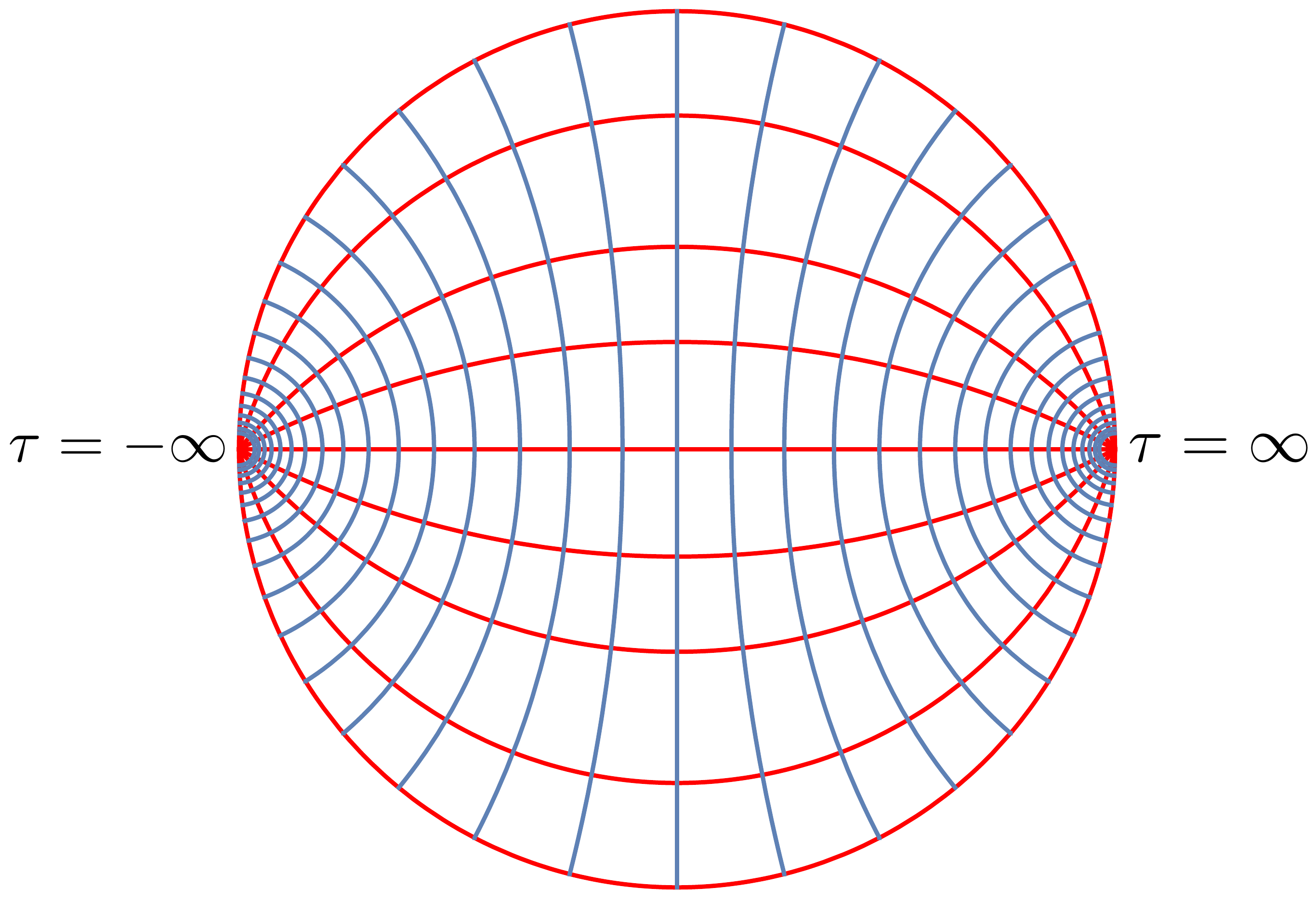}
\caption{Curves of constant $\tau$ (in blue) and constant $\rho$ (in red)
for AdS$_{2}$ stereographically projected to the unit disk (Poincaré disk). This shows how surfaces of constant $\tau$ converge to a boundary
bound when $\tau\to-\infty$. The cartesian coordinates in the plane
of the figure are given by $\vec{w}=\frac{(\cosh\rho\sinh\tau,\sinh\rho)}{1+\cosh\rho\cosh\tau}$
which puts the AdS$_{2}$ metric in the form $ds^{2}=\frac{4d\vec{w}^{2}}{1-\vec{w}^{2}}$.\label{fig:Poincare Disk}}
\end{centering}
\end{figure}

The map from states to boundary operators can be established using
global time translation invariance, 
\begin{align}
&\left\langle 0\right|\dots\hat{\phi}(\tau_{3},\rho_{3},\Omega_{3})\left|\psi(0)\right\rangle \\
=&\lim_{\tau\to-\infty}\left\langle 0\right|\dots\hat{\phi}(\tau_{3},\rho_{3},\Omega_{3})e^{\tau D}\left|\psi(\tau)\right\rangle \equiv\left\langle \dots\phi(X_{3})\mathcal{O}_{\psi}(P)\right\rangle \nonumber
\end{align}
where $|\psi(\tau)\rangle=e^{-\tau D}|\psi\rangle$ and $P^{A}=\left(\frac{1}{2},0,\frac{1}{2}\right)$
is again the boundary point defined by $\tau\to-\infty$ in global
coordinates. The idea is that $|\psi(\tau)\rangle$ prepares a boundary
condition for the path integral on a surface of constant $\tau$ and
this surface converges to a small cap around the boundary point $P^{A}=\left(\frac{1}{2},0,\frac{1}{2}\right)$
when $\tau\to-\infty$. This is depicted in figure \ref{fig:Poincare Disk}.

The Hilbert space of the bulk theory can be decomposed in irreducible
representations of the isometry group $SO(d+1,1)$. These are the
highest weight representations of the conformal group, labelled by
the scaling dimension and $SO(d)$ irrep of the the primary state.
Therefore, the CFT conformal block decomposition of correlators follows
from the partial wave decomposition in AdS, i.e. the decomposition
in intermediate eigenstates of the Hamiltonian organized in irreps
of the isometry group $SO(d+1,1)$. For example, the conformal block
decomposition of the disconnected part of the four-point function,
\be
\left\langle \mathcal{O}(P_{1})\dots\mathcal{O}(P_{4})\right\rangle =
\frac{1}{\left( P_{12}  P_{34}\right)^{\Delta} }+
\frac{1}{\left( P_{13}  P_{24}\right)^{\Delta} }+
\frac{1}{\left( P_{14}  P_{23}\right)^{\Delta} } \ ,
\ee
where $P_{ij} =-2P_i \cdot P_j$,
 is given by a sum of conformal blocks associated with the vacuum
and two-particle intermediate states 
\be
\left\langle \mathcal{O}(P_{1})\dots\mathcal{O}(P_{4})\right\rangle =G_{0,0}(P_{1},\dots,P_{4})+\sum_{{l=0\atop even}}^{\infty}\sum_{n=0}^{\infty}c_{n,l}\, G_{2\Delta+2n+l,l}(P_{1},\dots,P_{4})\ .
\nonumber
\ee
\begin{exercise} 
Check this statement in $d=2$ using the formula
\cite{hep-th/0011040}
\be
G_{E,l}(P_{1},P_{2},P_{3},P_{4})=\frac{k(E+l,z)k(E-l,\bar{z})+k(E-l,z)k(E+l,\bar{z})}{\left(-2P_{1}\cdot P_{2}\right)^{\Delta}\left(-2P_{3}\cdot P_{4}\right)^{\Delta}\left(1+\delta_{l,0}\right)}
\ee
where 
\be
k(2\beta,z)=(-z)^{\beta}\ _{2}F_{1}(\beta,\beta,2\beta,z)\ .
\ee
Determine the coefficients $c_{n,l}$ for $n\le 1$ by matching the Taylor series expansion around $z=\bar{z}=0$. {\bf Extra:} using a computer you can compute many coefficients and guess the general formula.
\end{exercise}

\subsubsection{Generating function}

There is an equivalent way of defining CFT correlation functions from
QFT in AdS. We introduce the generating function 
\be
W\left[\phi_{b}\right]=\left\langle e^{\int_{\partial AdS}dP\phi_{b}(P)\mathcal{O}(P)}\right\rangle \ ,
\ee
where the integral over $\partial AdS$ denotes an integral over a
chosen section of the null cone in $\mathbb{R}^{d+1,1}$ with its
induced measure. We impose that the source obeys $\phi_{b}(\lambda P)=\lambda^{\Delta-d}\phi_{b}(P)$
so that the integral is invariant under a change of section, i.e.
conformal invariant. For example, in the Poincaré section the integral
reduces to $\int d^{d}x\phi_{b}(x)\mathcal{O}(x)$. Correlation functions
are easily obtained with functional derivatives
\be
\left\langle \mathcal{O}(P_{1})\dots\mathcal{O}(P_{n})\right\rangle =\left.\frac{\delta}{\delta\phi_{b}(P_{1})}\dots\frac{\delta}{\delta\phi_{b}(P_{n})}W\left[\phi_{b}\right]\right|_{\phi_{b}=0}\ .
\ee
If we set the generating function to be equal to the path integral
over the field $\phi$ in AdS 
\begin{equation}
W\left[\phi_{b}\right]=\frac{\int_{\phi\to\phi_{b}}\left[d\phi\right]e^{-S[\phi]}}{\int_{\phi\to0}\left[d\phi\right]e^{-S[\phi]}}\ ,\label{eq:AdSgeneratingfunctionW}
\end{equation}
with the boundary condition that it approaches the source $\phi_{b}$
at the boundary,
\begin{equation}
\lim_{\lambda\to\infty}\lambda^{d-\Delta}\phi(X=\lambda P+\dots)=\frac{1}{2\Delta-d}\frac{1}{\sqrt{C_{\Delta}}}\phi_{b}(P)\ ,\label{eq:AdSphiBC}
\end{equation}
then we recover the correlation functions of $\mathcal{O}$ defined
above as limits of the correlation functions of $\phi$. 

For a quadratic bulk action, tha ratio of path intagrals in (\ref{eq:AdSgeneratingfunctionW})
is given $e^{-S}$ computed on the classical solution obeying the
required boundary conditions. A natural guess for this solution is 
\begin{equation}
\phi(X)=\sqrt{C_{\Delta}}\int_{\partial AdS}dP\frac{\phi_{b}(P)}{(-2P\cdot X)^{\Delta}}\ .\label{eq:phiclassicalfromphib}
\end{equation}
This automatically solves the AdS equation of motion $\nabla^{2}\phi=m^{2}\phi$,
because it is an homogeneous function of weight $-\Delta$ and it
obeys $\partial_{A}\partial^{A}\phi=0$ in the embedding space (see
equations (\ref{eq:CasimirEmbedding}) and (\ref{eq:CasimirAdS})).
To see that it also obeys the boundary condition (\ref{eq:AdSphiBC})
it is convenient to use the Poincaré section.
\begin{exercise} 
In the Poincaré section (\ref{eq:boundAdSPoincareEmb})
and using Poincaré coordinates (\ref{eq:AdSPoincareEmb}), formula
(\ref{eq:phiclassicalfromphib}) reads
\begin{equation}
\phi\left(z,x\right)=\sqrt{C_{\Delta}}\int d^{d}y\frac{z^{\Delta}\phi_{b}(y)}{\left(z^{2}+(x-y)^{2}\right)^{\Delta}}\label{eq:phifromphibPoincare}
\end{equation}
and (\ref{eq:AdSphiBC}) reads 
\begin{equation}
\lim_{z\to0}z^{\Delta-d}\phi(z,x)=\frac{1}{2\Delta-d}\frac{1}{\sqrt{C_{\Delta}}}\phi_{b}(x)\ .\label{eq:AdSPhiBCPoincare}
\end{equation}
Show that (\ref{eq:AdSPhiBCPoincare}) follows from (\ref{eq:phifromphibPoincare}).
You can assume $2\Delta>d$.
\end{exercise}

The cubic term $\frac{1}{3!}g\phi^{3}$ in the action will lead to
(calculable) corrections of order $g$ in the classical solution (\ref{eq:phiclassicalfromphib}).
To determine the generating function $W[\phi_{b}]$ in the classical
limit we just have to compute the value of the bulk action (\ref{eq:actionAdSscalar})
on the classical solution. However, before doing that, we have to
address a small subtlety. We need to add a boundary term to the action
(\ref{eq:actionAdSscalar}) in order to have a well posed variational
problem.
\begin{exercise}
The coefficient $\beta$ should be chosen such that
the quadratic action %
\footnote{Here $w$ stands for a generic coordinate in AdS and the index $\alpha$
runs over the $d+1$ dimensions of AdS. %
}

\be
S_{2}=\int_{AdS}dw\sqrt{G}\left[\frac{1}{2}\left(\nabla\phi\right)^{2}+\frac{1}{2}m^{2}\phi^{2}\right]+\beta\int_{AdS}dw\sqrt{G}\,\nabla_{\alpha}\left(\phi\nabla^{\alpha}\phi\right)
\ee
is stationary around a classical solution obeying (\ref{eq:AdSPhiBCPoincare})
for any variation $\delta\phi$ that preserves the boundary condition,
i.e.
\be
\delta\phi(z,x)=z^{\Delta}\left[f(x)+O(z)\right]\ .
\ee
Show that $\beta=\frac{\Delta-d}{d}$ and that the on-shell action
is given by a boundary term
\be
S_{2}=\frac{2\Delta-d}{2d}\int_{AdS}dw\sqrt{g}\,\nabla_{\alpha}\left(\phi\nabla^{\alpha}\phi\right)\ .
\ee
Finally, show that for the classical solution (\ref{eq:phifromphibPoincare})
this action is given by %
\footnote{This integral is divergent if the source $\phi_{b}$ is a smooth function
and $\Delta>\frac{d}{2}$. The divergence comes from the short distance
limit $y_{1}\to y_{2}$ and does not affect the value of correlation
functions at separate points. Notice that a small value of $z>0$
provides a UV regulator. %
}
\be
S_{2}=-\frac{1}{2}\int d^{d}y_{1}d^{d}y_{2}\phi_{b}(y_{1})\phi_{b}(y_{2})K(y_{1},y_{2})\ ,
\ee
where
\begin{align}
K(y_{1},y_{2})&=C_{\Delta}\frac{2\Delta-d}{d}
\lim_{z\to0}\int \frac{d^{d}x}{z^{d-1}}
\frac{z^{\Delta}}{\left(z^{2}+(x-y_{1})^{2}\right)^{\Delta}}\partial_{z}\frac{z^{\Delta}}{\left(z^{2}+(x-y_{2})^{2}\right)^{\Delta}}\nonumber\\
&=\frac{1}{(y_{1}-y_{2})^{2\Delta}}
\end{align}
is the CFT two point function (\ref{eq:CFT2ptfromAdS}).
\end{exercise}
\begin{exercise}
Using $\phi=\phi_{0}+O(g)$ with $\phi_{0}$ given
by (\ref{eq:phiclassicalfromphib}), show that the complete on-shell
action is given by
\be
S=-\frac{1}{2}\int d^{d}y_{1}d^{d}y_{2}\phi_{b}(y_{1})\phi_{b}(y_{2})K(y_{1},y_{2})+\frac{1}{3!}g\int_{AdS}dX\left[\phi_{0}(X)\right]^{3}+O(g^{2})\ , \nonumber
\ee
and that this leads to the three-point function (\ref{eq:CFT3ptFromAdS}). 
{\bf Extra:} Compute the terms of $O(g^{2})$ in the on-shell action.
\end{exercise}

We have seen that QFT on an AdS background naturally defines conformal 
correlation functions living on the boundary of AdS. Moreover, we saw that a weakly coupled theory
in AdS gives rise to factorization of CFT correlators like in a large
$N$ expansion. However, there is one missing ingredient to obtain
a full-fledged CFT: a stress-energy tensor. In the next section, we
will see that this requires dynamical gravity in AdS. The next exercise
also shows that a free QFT in AdS$_{d+1}$ can not be dual to a local
CFT$_{d}$. 
\begin{exercise}
Compute the free-energy of a gas of free scalar particles
in AdS. Since particles are free and bosonic one can create multi-particle
states by populating each single particle state an arbitrary number
of times. That means that the total partition function is a product
over all single particle states and it is entirely determined by the
single particle partition function. More precisely, show that 
\begin{align}
&F=-T\log Z=-T\log\prod_{\psi_{sp}}\left(\sum_{k=0}^{\infty}q^{kE_{\psi_{sp}}}\right)=-T\sum_{n=1}^{\infty}\frac{1}{n}Z_{1}\left(q^{n}\right)\ ,\\
&Z_{1}(q)=\sum_{\psi_{sp}}q^{E_{\psi_{sp}}}=\frac{q^{\Delta}}{(1-q)^{d}}\ ,
\end{align}
where $q=e^{-\frac{1}{RT}}$ and we have used the single-particle spectrum
of the hamiltonian $D=-\frac{\partial}{\partial\tau}$ of AdS in global
coordinates. Show that
\be
F\approx-\zeta(d+1)R^dT^{d+1}
\ee
in the high temperature regime and compute the entropy using the thermodynamic
relation $S=-\frac{\partial F}{\partial T}$. Compare this result
with the expectation 
\be
S\sim(RT)^{d-1}\ ,
\ee
for the high temperature behaviour of the entropy of a CFT on a sphere
$S^{d-1}$ of radius $R$. See section 4.3 of reference \cite{arXiv:1101.4163}
for more details.
\end{exercise}

\subsection{Gravity with AdS boundary conditions\label{sub:GravityAdS}}

Consider general relativity in the presence of a negative cosmological
constant
\begin{equation}
I[G]=\frac{1}{\ell_{P}^{d-1}}\int d^{d+1}w\sqrt{G}\left[\mathcal{R}-2\Lambda\right]\ .\label{eq:GRactionAdS}
\end{equation}
The AdS geometry 
\be
ds^{2}=R^{2}\frac{dz^{2}+dx_{\mu}dx^{\mu}}{z^{2}}\ ,
\ee
is a maximally symmetric classical solution with $\Lambda=-\frac{d(d-1)}{2R^{2}}$.
When the AdS radius $R$ is much larger than the Planck length $\ell_{P}$
the metric fluctuations are weakly coupled and form an approximate
Fock space of  graviton states. One can compute the single graviton
states and verify that they are in one-to-one correspondence with
the CFT stress-tensor operator and its descendants (with AdS energies
matching scaling dimensions). One can also obtain CFT correlation
functions of the stress-energy tensor using Witten diagrams in AdS.
The new ingredients are the bulk to boundary and bulk to bulk graviton
propagators \cite{hep-th/9804083, hep-th/9807097, hep-th/9902042, hep-th/9903196, arXiv:1404.5625}.

In the gravitational context, it is nicer to use the partition function
formulation
\begin{equation}
Z[g_{\mu\nu},\phi_{b}]=\int_{{G\to g\atop \phi\to\phi_{b}}}\left[dG\right]\left[d\phi\right]e^{-I[G,\phi]}\label{eq:AdSgeneratingfunctionmetric}
\end{equation}
where 
\be
I[G,\phi]=\frac{1}{\ell_{P}^{d-1}}\int d^{d+1}w\sqrt{G}\left[\mathcal{R}-2\Lambda+\frac{1}{2}\left(\nabla\phi\right)^{2}+\frac{1}{2}m^{2}\phi^{2}\right]
\ee
and the boundary condition are 
\begin{eqnarray}
ds^{2} & = & G_{\alpha\beta}dw^{\alpha}dw^{\beta}=R^{2}\frac{dz^{2}+dx^{\mu}dx^{\nu}\left[g_{\mu\nu}(x)+O(z)\right]}{z^{2}}\ ,\label{eq:BCAdSmetric}\\
\phi & = & \frac{z^{d-\Delta}}{2\Delta-d}\left[\phi_{b}(x)+O(z)\right]\ .\nonumber 
\end{eqnarray}
By construction the partition function is invariant under diffeomorphisms
of the boundary metric $g_{\mu\nu}$. Therefore, this definition implies
the Ward identity (\ref{eq:DiffWardIdentity}). The generating function
is also invariant under Weyl transformations
\be
Z\left[\Omega^{2}g_{\mu\nu},\Omega^{\Delta-d}\phi_{b}\right]=Z\left[g_{\mu\nu},\phi_{b}\right]\qquad(naive)
\ee
This follows from the fact that the boundary condition
\begin{eqnarray}
ds^{2} & = & R^{2}\frac{dz^{2}+dx^{\mu}dx^{\nu}\left[\Omega^{2}(x)g_{\mu\nu}(x)+O(z)\right]}{z^{2}}\label{eq:BCAdSmetricWeyl}\\
\phi & = & \frac{z^{d-\Delta}}{2\Delta-d}\left[\Omega^{\Delta-d}(x)\phi_{b}(x)+O(z)\right]\nonumber 
\end{eqnarray}
can be mapped to (\ref{eq:BCAdSmetric}) by the following coordinate
transformation
\begin{eqnarray}
z & \to & z\,\Omega-\frac{1}{4}z^{3}\Omega\left(\partial_{\mu}\log\Omega\right)^{2}+O(z^{5})\label{eq:coordtransWeylAds}\\
x^{\mu} & \to & x^{\mu}-\frac{1}{2}z^{2}\partial^{\mu}\log\Omega+O(z^{4})\nonumber 
\end{eqnarray}
where indices are raised and contracted using the metric $g_{\mu\nu}$
and its inverse. In other words, a bulk geometry that satisfies (\ref{eq:BCAdSmetric})
also satisfies (\ref{eq:BCAdSmetricWeyl}) with an appropriate choice
of coordinates. If the partition function (\ref{eq:AdSgeneratingfunctionmetric})
was a finite quantity this would be the end of the story. However,
even in the classical limit, where $Z\approx e^{-I}$, the partition
function needs to be regulated. The divergences originate from the
$z\to0$ region and can be regulated by cutting off the bulk integrals
at $z=\epsilon$ (as it happened for the scalar case discussed above).
Since the coordinate transformation (\ref{eq:coordtransWeylAds})
does not preserve the cutoff, the regulated partition function is
not obviously Weyl invariant. This has been studied in great detail
in the context of holographic renormalization \cite{hep-th/9902121, hep-th/0209067}.
In particular, it leads to the Weyl anomaly $g^{\mu\nu}T_{\mu\nu}\neq0$
when $d$ is even. The crucial point is that this is a UV effect that
does not affect the connected correlation functions of operators at
separate points.  
In particular, the integrated form (\ref{eq:WardIdentityFlux})=(\ref{eq:WardIndentityTransf})
of the conformal Ward identity is valid.

We do not now how to define the quantum gravity path integral in (\ref{eq:AdSgeneratingfunctionmetric}).
The best we can do is a semiclassical expansion when $\ell_{P}\ll R$.
This semiclassical expansion gives rise to connected correlators of
the stress tensor $T_{\mu\nu}$ that scale as 
\be
\left\langle T_{\mu_{1}\nu_{1}}(x_{1})\dots T_{\mu_{n}\nu_{n}}(x_{n})\right\rangle _{c}\sim\left(\frac{R}{\ell_{P}}\right)^{d-1}\ .
\ee
 This is exactly the scaling (\ref{eq:TTTTTlargeNscaling}) we found
from large $N$ factorization if we identify $N^{2}\sim\left(\frac{R}{\ell_{P}}\right)^{d-1}$.
This suggests that CFTs related to semiclassical Einstein gravity
in AdS, should have a large number of local degrees of freedom. This
can be made more precise. The two-point function of the stress tensor
in a CFT is given by
\be
\left\langle T_{\mu\nu}(x)T_{\sigma\rho}(0)\right\rangle =\frac{C_{T}}{S_{d}^{2}}\frac{1}{x{}^{2d}}\left[\frac{1}{2}I_{\mu\sigma}I_{\nu\rho}+\frac{1}{2}I_{\mu\rho}I_{\nu\sigma}-\frac{1}{d}\delta_{\mu\nu}\delta_{\sigma\rho}\right]\ ,
\ee
where $S_{d}=\frac{2\pi^{d/2}}{\Gamma(d/2)}$ is the volume of a $(d-1)$-dimensional
unit sphere and 
\be
I_{\mu\nu}=\delta_{\mu\nu}-2\frac{x_{\mu}x_{\nu}}{x^{2}}\ .
\ee
The constant $C_{T}$ provides an (approximate) measure of the number of degrees
of freedom.\footnote{However, for $d>2$, $C_T$ is not a $c$-function that always decreases under  Renormalization Group flow.}
For instance, for $n_{\varphi}$ free scalar fields and
$n_{\psi}$ free Dirac fields we find \cite{hep-th/9307010}%
%\footnote{A free gauge field in $d=4$ contributes 16 to $C_{T}$.}
\be
C_{T}=n_{\varphi}\frac{d}{d-1}+n_{\psi}2^{\left[\frac{d}{2}\right]-1}d\ ,
\ee
where $\left[x\right]$ is the integer part of $x$.
If the CFT is described by Einstein gravity in AdS, we find \cite{hep-th/9804083}
\be
C_{T}=8\frac{d+1}{d-1}\frac{\pi^{\frac{d}{2}}\Gamma(d+1)}{\Gamma^{3}\left(\frac{d}{2}\right)}\frac{R^{d-1}}{\ell_{P}^{d-1}}\ ,
\ee
which shows that the CFT dual of a semiclassical gravitational theory
with $R\gg\ell_{P}$, must have a very large number of degrees of
freedom.

In summary, semiclassical gravity with AdS boundary conditions gives
rise to a set of correlation functions that have all the properties
(conformal invariance, Ward identities, large $N$ factorization)
expected for the correlation functions of the stress tensor of a large
$N$ CFT. Therefore, it is natural to ask if a CFT with finite $N$
is a quantum theory of gravity.

\section{The AdS/CFT Correspondence \label{sec:AdSCFT}}

\subsection{Quantum Gravity as CFT \label{sec:QGinAdS}}

What is quantum gravity? The most conservative answer is a standard
quantum mechanical theory whose low energy dynamics around a weakly
curved background is well described by general relativity (or some
other theory with a dynamical metric). This viewpoint is particularly
useful with asymptotically AdS boundary conditions. In this case,
we can view the AdS geometry with a radius much larger than the Planck
length as a background for excitations (gravitons) that are weakly
coupled at low energies. Therefore, we just need to find a quantum
system that reproduces the dynamics of low energy gravitons in a large
AdS. In fact, we should be more precise about the word ``reproduces''.
We should define observables in quantum gravity that our quantum system
must reproduce. This is not so easy because the spacetime geometry
is dynamical and we can not define local operators. In fact, the only
well defined observables are defined at the (conformal) boundary like
the partition function (\ref{eq:AdSgeneratingfunctionmetric}) and
the associated correlation functions obtained by taking functional
derivatives. But in the previous section we saw that these observables
have all the properties expected for the correlation functions of
a large $N$ CFT. Thus, quantum gravity with AdS boundary conditions
is equivalent to a CFT.

There are many CFTs and not all of them are useful theories of quantum
gravity. Firstly, it is convenient to consider a family of CFTs labeled
by $N$, so that we can match the bulk semiclassical expansion using
$N^{2}\sim\left(\frac{R}{\ell_{P}}\right)^{d-1}$. In the large $N$
limit, every CFT single-trace primary operator of scaling dimension
$\Delta$ gives rise to a weakly coupled field in AdS with mass $m\sim\Delta/R$.
Therefore, if are looking for a UV completion of pure gravity in AdS
without any other low energy fields, then we need to find a CFT where
all single-trace operators have parametrically large dimension, except
the stress tensor. This requires strong coupling and seems rather
hard to achieve. Notice that a weakly coupled CFT with gauge group
$SU(N)$ and fields in the adjoint representation has an infinite
number of primary single-trace operators with order 1 scaling dimension.
It is natural to conjecture that large $N$ factorization and correct
spectrum of single-trace operators are sufficient conditions for a
CFT to provide a UV completion of General Relativity (GR) \cite{arXiv:0907.0151}.
However, this is not obvious because we still have to check if the
CFT correlation functions of $T_{\mu\nu}$ match the prediction from
GR in AdS. For example, the stress tensor three-point function is
fixed by conformal symmetry to be a linear combination of 3 independent
conformal invariant structures. 
\footnote{Here we are assuming $d\ge4$. For $d=3$ there are only 2 independent
structures.} On the other hand, the action (\ref{eq:GRactionAdS}) predicts a
specific linear combination. It is not obvious that all large $N$
CFTs with the correct spectrum will automatically give rise to the
same three-point function. There has been some recent progress in
this respect. The authors of \cite{arXiv:1407.5597} used causality
to show that this is the case. Unfortunately, their argument uses
the bulk theory and can not be formulated entirely in CFT language.
In any case, this is just the three-point fuction and GR predicts
the leading large $N$ behaviour of all $n$-point functions. It is
an important open problem to prove the following conjecture:

\emph{Any large N CFT where all single-trace operators, except the stress tensor, have parametrically large scaling dimensions, has the stress tensor correlation functions predicted by General Relativity in AdS.}

Perhaps the most pressing question is if such CFTs exist at all. At
the moment, we do not know the answer to this question but in the
next section we will discuss closely related examples that are realized
in the context of string theory.

If some CFTs are theories of quantum gravity, it is natural to ask
if there are other CFT observables with a nice gravitational interpretation.
One interesting example that will be extensively discussed in this
school is the entanglement entropy of a subsystem. In section \ref{sub:Finite-Temperature},
we will discuss how CFT thermodynamics compares with black hole thermodynamics
in AdS. In addition, in section \ref{sub:Applications} we will give
several examples of QFT phenomena that have beautiful geometric meaning
in the holographic dual.

\subsection{String Theory}

The logical flow presented above is very different from the historical
route that led to the AdS/CFT correspondence. Moreover, from what
we said so far AdS/CFT looks like a very abstract construction without
any concrete examples of CFTs that have simple gravitational duals.
If this was the full story probably I would not be writing these lecture notes.
The problem is that we have stated properties that we want for our
CFTs but we have said nothing about how to construct these CFTs besides
the fact that they should be strongly coupled and obey large $N$
factorization. Remarkably, string theory provides a ``method'' to
find explicit examples of CFTs and their dual gravitational theories. 

The basic idea is to consider the low energy description of D-brane
systems from the perspective of open and closed strings. Let us illustrate
the argument by quickly summarizing the prototypical example of AdS/CFT
\cite{hep-th/9711200}. Consider $N$ coincident D3-branes of type
IIB string theory in 10 dimensional Minkowski spacetime. Closed strings
propagating in 10 dimensions can interact with the D3-branes. This
interaction can be described in two equivalent ways: 

(a) D3-branes can be defined as a submanifold where open strings can
end. This means that a closed string interacts with the D3-branes
by breaking the string loop into an open string with endpoints attached
to the D3-branes. 

(b) D3-branes can be defined as solitons of closed string theory. In
other words, they create a non-trivial curved background where closed
strings propagate.

\begin{figure}
\begin{centering}
\includegraphics[width=0.6\textwidth]{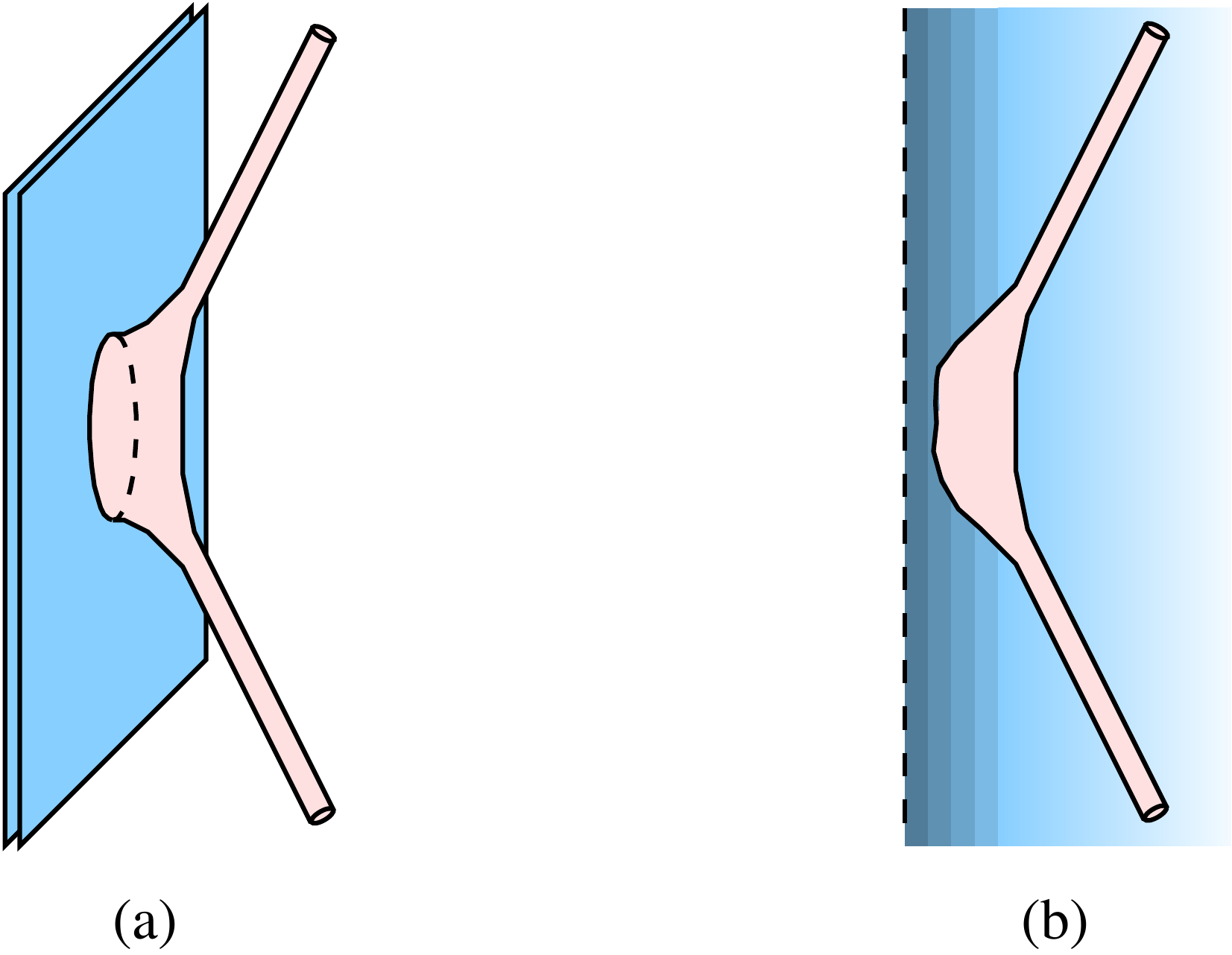}
\par\end{centering}

\protect\caption{\label{fig:OpenClosedDuality}(a) Closed string scattering off branes
in flat space. (b) Closed string propagating in a curved background.}

\end{figure}

These two alternatives are depicted in figure \ref{fig:OpenClosedDuality}.
Their equivalence is called open/closed duality. The AdS/CFT correspondence
follows from the low-energy limit of open/closed duality. We implement
this low-energy limit by taking the string length $\ell_{s}\to0$,
keeping the string coupling $g_{s}$, the number of branes $N$ and
the energy fixed. In description (a), the low energy excitations of
the system form two decoupled sectors: massless closed strings propagating
in 10 dimensional Minkoski spacetime and massless open strings attached
to the D3-branes, which at low energies are well described by $\mathcal{N}=4$
Supersymmetric Yang-Mills (SYM) with gauge group $SU(N)$. In description
(b), the massless closed strings propagate in the following geometry
\be
ds^{2}=\frac{1}{\sqrt{H(r)}}\eta_{\mu\nu}dx^{\mu}dx^{\nu}+\sqrt{H(r)}\left[dr^{2}+r^{2}d\Omega_{5}^{2}\right]\ ,
\ee
where $\eta_{\mu\nu}$ is the metric of the 4 dimensional Minkowski
space along the branes and 
\be
H(r)=1+\frac{R^{4}}{r^{4}}\ ,\qquad R^{4}=4\pi g_{s}N\ell_{s}^{4}\ .
\ee
Naively, the limit $\ell_{s}\to0$ just produces 10 dimensional Minkowski
spacetime. However, one has to be careful with the region close to
the branes at $r=0$. A local high energy excitation in this region
will be very redshifted from the point of view of the observer at
infinity. In order to determine the correct low-energy limit in the
region around $r=0$ we introduce a new coordinate $z=R^{2}/r$, which
we keep fixed as $\ell_{s}\to0$. This leads to
\be
ds^{2}=R^{2}\frac{dz^{2}+\eta_{\mu\nu}dx^{\mu}dx^{\nu}}{z^{2}}+R^{2}d\Omega_{5}^{2}\ ,
\ee
which is the metric of AdS$_{5}\times S^{5}$ both with radius $R$.
Therefore, description (b) also leads to 2 decoupled sectors of low
energy excitations: massless closed strings in 10D and full type IIB
string theory on AdS$_{5}\times S^{5}$. This led Maldacena to conjecture
that

\begin{center}
\begin{tabular}{ccc}
$SU(N)$ SYM & \multirow{2}{*}{$\Leftrightarrow$} & IIB strings on AdS$_{5}\times S^{5}$ \tabularnewline
\multirow{1}{*}{$g_{YM}^{2}=4\pi g_{s}$} &  & $\frac{R^{4}}{\ell_{s}^{4}}=g_{YM}^{2}N\equiv\lambda$\tabularnewline
\end{tabular}
\par\end{center}

SYM is conformal for any value of $N$ and the coupling constant $g_{YM}^{2}$.
The lagrangian of the theory involves the field strength 
\be
F_{\mu\nu}=\partial_{\mu}A_{\nu}-\partial_{\nu}A_{\mu}-i\left[A_{\mu},A_{\nu}\right]\ ,
\ee
6 scalars fields $\Phi^{m}$ and 4 Weyl fermions $\Psi^{a}$, which
are all valued in the adjoint representation of $SU(N)$. The lagrangian
is given by
\begin{align}
&\frac{1}{g_{YM}^{2}}{\rm Tr}\left[\frac{1}{4}F^{\mu\nu}F_{\mu\nu}+\frac{1}{2}\left(D^{\mu}\Phi^{m}\right)^{2}+\bar{\Psi}^{a}\sigma^{\mu}D_{\mu}\Psi_{a}\right.\\
&\qquad\qquad\left.-\frac{1}{4}\left[\Phi^{m},\Phi^{n}\right]^{2}-C_{m}^{ab}\Psi_{a}\left[\Phi^{m},\Psi_{b}\right]-\bar{C}_{mab}\bar{\Psi}^{a}\left[\Phi^{m},\bar{\Psi}^{b}\right]\right]\ , \nonumber
\end{align}
where $D_{\mu}$ is the gauge covariant derivative and $C_{m}^{ab}$
and $\bar{C}_{mab}$ are constants fixed by the $SO(6)=SU(4)$ global symmetry
of the theory. Notice that the isometry group of AdS$_{5}\times S^{5}$
is $SO(5,1)\times SO(6)$, which matches precisely the bosonic symmetries
of SYM: conformal group $\times$ global $SO(6)$. There are many
interesting things to say about SYM. In some sense, SYM is the simplest
interacting QFT in 4 dimensions \cite{ArkaniHamed:2008gz}. However, this is not the focus of
these lectures and we refer the reader to the numerous existing reviews
about SYM \cite{hep-th/0201253, arXiv:1012.3982}.

The remarkable conjecture of Maldacena has been extensively tested
since it was first proposed in 1997 \cite{hep-th/9711200}. To test
this conjecture one has to be able to compute the same observable
on both sides of the duality. This is usually a very difficult task.
On the SYM side, the regime accessible to perturbation theory is $g_{YM}^{2}N\ll1$.
This implies $g_{s}\ll1$, which on the string theory side suppresses
string loops. However, it also implies that the AdS radius of curvature
$R$ is much smaller than the string length $\ell_{s}$. This means
that the string worldsheet theory is very strongly coupled. In fact,
the easy regime on the string theory side is $g_{s}\ll1$ and $R\gg\ell_{s}$,
so that (locally) strings propagate in an approximately flat space.
Thus, directly testing the conjecture is a formidable task. There
are three situations where a direct check can be made analitycally. 

The first situation arises when some observable is independent of
the coupling constant. In this case, one can compute it at weak coupling
$\lambda\ll1$ using the field theory description and at strong coupling
$\lambda\gg1$ using the string theory description. Usually this involves
completely different techniques but in the end the results agree.
Due to the large supersymmetry of SYM there are many observables that
do not depend on the coupling constant. Notable examples include the
scaling dimensions of half BPS single-trace operators and their three-point
functions \cite{hep-th/9806074}.

The second situation involves observables that depend on the coupling
constant $\lambda$ but preserve enough supersymmetry that can be
computed at any value of $\lambda$ using a technique called localization.
Important examples of this type are the sphere partition function
and the expectation value of circular Wilson loops \cite{arXiv:0712.2824 , hep-th/0003055}.

Finally, the third situation follows from the conjectured integrability
of SYM in the planar limit. Assuming integrability one can compute
the scaling dimension of non-protected single-trace operators at any
value of $\lambda$ and match this result with SYM perturbative calculations
for $\lambda\ll1$ and with weakly coupled string theory for $\lambda\gg1$
(see figure 1 from \cite{arXiv:0906.4240}). Planar scattering amplitudes
an three-point functions of single-trace operators can also be computed
using integrability \cite{arXiv:1303.1396, arXiv:1505.06745}.

There are also numerical tests of the gauge/gravity duality. 
The most impressive study in this context, was the Monte-Carlo simulation  
of the BFSS matrix model \cite{hep-th/9610043} at finite temperature that reproduced the predictions from its dual 
black hole geometry \cite{arXiv:0706.3518, arXiv:0707.4454, arXiv:0803.4273, arXiv:0811.3102, arXiv:0909.4947, arXiv:1108.5153, arXiv:1311.5607}. 

How does the Maldacena conjecture fit into the general discussion
of the previous sections? One important novelty is the presence of
a large internal sphere on the gravitational side. We can perform
a Kaluza-Klein reduction on $S^{5}$ and obtain an effective action
for AdS$_{5}$
\be
\frac{1}{(2\pi)^{7}\ell_{s}^{8}}\int d^{10}x\sqrt{g_{10}}e^{-2\Phi}\left[\mathcal{R}_{10}+\dots\right]\to\frac{R^{5}}{8(2\pi)^{4}g_{s}^{2}\ell_{s}^{8}}\int d^{5}x\sqrt{g_{5}}\left[\mathcal{R}_{5}+\dots\right]\ .
\nonumber
\ee
This allows us to identify the 5 dimensional Planck length 
\be
\ell_{P}^{3}=\frac{8(2\pi)^{4}g_{s}^{2}\ell_{s}^{8}}{R^{5}}
\ee
and verify the general prediction $N^{2}\sim R^{3}/\ell_{P}^{3}$
. Remarkably, at strong coupling $\lambda\gg1$ all single-trace non-protected
operators of SYM have parametrically large scaling dimensions. This
is simple to understand from the string point of view. Massive string
states have masses $m\sim1/\ell_{s}$. But we saw in the previous
sections that the dual operator to an AdS field of mass $m$ has a
scaling dimension $\Delta\sim mR\sim R/\ell_{s} \sim\lambda^{\frac{1}{4}}$. The only
CFT operators that have small scaling dimension for $\lambda\gg1$
are dual to massless string states that constitute the fields of type
IIB supergravity (SUGRA). Therefore, one can say that SYM (with $N\gg\lambda\gg1$)
provides a UV completion of IIB SUGRA with AdS$_{5}\times S^{5}$
boundary conditions. 

String theory provides more concrete examples of AdS/CFT dual pairs.
These examples usually involve SCFTs (or closely related non-supersymmetry
theories). This is surprising because SUSY played no role in our general
discussion. At the moment, it is not known if SUSY is an essential
ingredient of AdS/CFT or if it is only a useful property that simplifies
the calculations. The latter seems more likely but notice that SUSY
might be essential to stabilize very strong coupling and allow the
phenomena of large scaling dimensions for almost all single-trace
operators. Another observation is that it turns out to be very difficult
to construct AdS duals with small internal spaces (for SYM we got
a 5-sphere with the same radius of AdS$_{5}$). It is an open problem
to find CFTs with gravity duals in less than 10 dimensions (see \cite{arXiv:0908.0756, arXiv:1412.6999}
for attempts in this direction).

Another interesting class of examples are the dualities between vector
models and Higher Spin Theories (HST) \cite{hep-th/0210114, hep-th/0205131}.
Consider for simplicity the free $O(n)$ model in 3 dimensions
\be
S=\int d^{3}x\sum_{i=1}^{n}\frac{1}{2}\partial_{\mu}\varphi^{i}\partial^{\mu}\varphi^{i}\ .
\ee
In this case, the analogue of single-trace operators are the $O(n)$
singlets $\mathcal{O}_{l}=\sum_{i}\varphi^{i}\partial_{\mu_{1}}\dots\partial_{\mu_{l}}\varphi^{i}$
with even spin $l$ and dimension $\Delta=1+l$. At large $n$, the
correlation functions of these operators factorize with $n$ playing
the role of $N^{2}$ in a $SU(N)$ gauge theory with adjoint fields.
The AdS dual of this CFT is a theory with one massless field for each
even spin. These theories are rather non-local and they can not be
defined in flat spacetime. Even if we introduce the relevant interaction
$\left(\varphi^{i}\varphi^{i}\right)^{2}$ and flow to the IR fixed
point (Wilson-Fisher fixed point), the operators $\mathcal{O}_{l}$
with $l>2$ get anomalous dimensions of order $\frac{1}{n}$ and therefore
the classical AdS theory still contains the same number of massless
higher spin fields. This duality has been extended to theories with
fermions and to theories where the global $O(n)$ symmetry is gauged
using Chern-Simons gauge fields. It is remarkable that HST in AdS
seems to have the correct structure to reproduce the CFT observables
that have been computed so far. Notice that in these examples of AdS/CFT
supersymmetry plays no role. However, it is unclear if the AdS description
is really useful in this case.%
\footnote{In practice  it was  very useful because it led to an intensive
study of Chern-Simons matter theories, which gave rise to the remarkable
conjecture of fermion/boson duality in 3 dimensions \cite{arXiv:1110.4386, arXiv:1207.4593}.%
} In practice, the large $n$ limit of these vector models is solvable
and the dual HST in AdS is rather complicated to work with even at
the classical level. There are also analogous models in AdS$_{3}$/CFT$_{2}$
duality \cite{arXiv:1207.6697}.

\subsection{Finite Temperature \label{sub:Finite-Temperature}}

In section \ref{sub:GravityAdS}, we argued that holographic CFTs
must have a large number of local degrees of freedom, using the two-point
function of the stress tensor. Another way of counting degrees of
freedom is to look at the entropy density when the system is put at
finite temperature. For a CFT in flat space and infinite volume, the
temperature dependence of the entropy density is fixed by dimensional
analysis because there is no other scale available,
\be
s=c_{s}T^{d-1}\ .
\ee
The constant $c_{s}$ is a physical measure of the number of degrees
of freedom. 

The gravitational dual of the system at finite temperature is a black
brane in asymptotically AdS space. The Euclidean metric is given by
\be
ds^{2}=\frac{R^{2}}{z^{2}}\left[\frac{dz^{2}}{1-(z/z_{H})^{d}}+\left(1-\frac{z^{d}}{z_{H}^{d}}\right)d\tau^{2}+\delta_{ij}dx^{i}dx^{j}\right]\ .
\ee
\begin{exercise} 
Show that in order to avoid a conical defect at the
horizon $z=z_{H}$, we need to identify Euclidean time $\tau$ with
period $\frac{4\pi z_{H}}{d}$. This fixes the Hawking temperature
$T=\frac{d}{4\pi z_{H}}$.
\end{exercise} 
The formula $T=\frac{d}{4\pi z_{H}}$ illustrates a general phenomena
in holography: high energy corresponds to the region close to the
boundary and low energy corresponds to the deep interior of the dual
geometry.

The entropy of the system is given by the Bekenstein-Hawking formula
\be
S=\frac{A_{H}}{4G_{N}}=\frac{4\pi}{\ell_{P}^{d-1}}\frac{R^{d-1}}{z_{H}^{d-1}}\int d^{d-1}x\qquad\Rightarrow\qquad c_{s}=\frac{(4\pi)^{d}}{d^{d-1}}\frac{R^{d-1}}{\ell_{P}^{d-1}}\ .
\ee
As expected $c_{s}$ is very large in the bulk classical limit $R\gg\ell_{P}$.
Interestingly, the ratio
\begin{equation}
\frac{c_{s}}{C_{T}}=\frac{\pi^{\frac{d}{2}}}{8}\left(\frac{4}{d}\right)^{d}\frac{d-1}{d+1}\frac{\Gamma^{3}\left(\frac{d}{2}\right)}{\Gamma(d)}\label{eq:csovercT}
\end{equation}
only depends on the spacetime dimension $d$ if the CFT has a classical
bulk dual \cite{arXiv:0801.2785}. It would be very nice to prove
that all large $N$ CFTs where all single-trace operators, except
the stress tensor, have parametrically large scaling dimensions, satisfy
(\ref{eq:csovercT}). Notice that (\ref{eq:csovercT}) is automatic
in $d=2$ because $C_{T}=2c$ and $c_{s}=\frac{\pi}{3}c$ are uniquely
fixed in terms of the central charge $c$. In planar SYM, $C_{T}=40N^{2}$
is independent of the 't Hooft coupling but $c_{s}$ varies with $\lambda$
(although not much, $c_{s}(\lambda=\infty)=\frac{3}{4}c_{s}(\lambda=0)$).
In this case, (\ref{eq:csovercT}) is only satisfied at strong coupling,
when all primary operators with spin greater than 2 have parametrically
large scaling dimensions. 

\begin{exercise}  
\label{ex:HPtransition}
Consider a CFT on a sphere of radius $L$ and at
temperature $T$. In this case, the entropy is a non-trivial function
of the dimensioless combination $LT$. Let us compute this function
assuming the CFT is well described by Einstein gravity with asymptotically
AdS boundary conditions. There are two possible bulk geometries that
asymptote to the Euclidean boundary $S^{1}\times S^{d-1}$. The first
is pure AdS 
\begin{equation}
ds^{2}=R^{2}\left[\frac{dr^{2}}{1+r^{2}}+\left(1+r^{2}\right)d\tau^{2}+r^{2}d\Omega_{d-1}^{2}\right]\label{eq:pureAdSglobal}
\end{equation}
 with Euclidean time periodically identified and the second is Schwarzschild-AdS
\begin{equation}
ds^{2}=R^{2}\left[\frac{dr^{2}}{f(r)}+f(r)d\tilde{\tau}^{2}+r^{2}d\Omega_{d-1}^{2}\right]\ ,\label{eq:AdSBH}
\end{equation}
where $f(r)=1+r^{2}-\frac{m}{r^{d-2}}$. At the boundary $r=r_{max}\gg1$,
both solutions should be conformal to $S^{1}\times S^{d-1}$ with
the correct radii. Show that this fixes the periodicities 
\be
\Delta\tau=\frac{1}{TL}\frac{r_{max}}{\sqrt{1+r_{max}^{2}}}\ ,\qquad\Delta\tilde{\tau}=\frac{1}{TL}\frac{r_{max}}{\sqrt{f(r_{max})}}\ .
\ee
Show also that regularity of the metric (\ref{eq:AdSBH}) implies
the periodicity
\be
\Delta\tilde{\tau}=\frac{4\pi}{f'(r_{H})}=\frac{4\pi}{r_{H}d+\frac{d-2}{r_{H}}}\ ,
\ee
where $r=r_{H}$ is the largest zero of $f(r)$. Notice that this
implies a minimal temperature for Schwarzschild black holes in AdS,
$T>\frac{\sqrt{d(d-2)}}{2\pi L}$.

Both (\ref{eq:pureAdSglobal}) and (\ref{eq:AdSBH}) are stationary
points of the Euclidean action (\ref{eq:GRactionAdS}). Therefore,
we must compute the value of the on-shell action in order to decide
which one dominates the path integral. Show that the difference of
the on-shell actions is given by
\begin{align}
I_{BH}-I_{AdS}&=-2S_{d}\frac{R^{d-1}}{\ell_{P}^{d-1}}\left[r_{max}^{d}\Delta\tau-\left(r_{max}^{d}-r_{H}^{d}\right)\Delta\tilde{\tau}\right]\\
&\longrightarrow S_{d}\frac{R^{d-1}}{\ell_{P}^{d-1}}\frac{1}{TL}r_{H}^{d-2}(1-r_{H}^{2})
\end{align}
where $S_{d}$ is the area of a unit $(d-1)$-dimensional sphere and
in the last step we took the limit $r_{max}\to\infty$. Conclude that
the black hole only dominates the bulk path integral when $r_{H}>1$,
which corresponds to $T>\frac{d-1}{2\pi L}$. This is the Hawking-Page
phase transition \cite{Hawking:1982dh}. It is natural to set the
free-energy of the AdS phase to zero because this phase corresponds
to a gas of gravitons around the AdS background whose free energy
does not scale with the large parameter $(R/\ell_{P})^{d-1}$. Therefore,
the free energy of the black hole phase is given by
\be
F_{BH}=\frac{1}{L}S_{d}\frac{R^{d-1}}{\ell_{P}^{d-1}}r_{H}^{d-2}(1-r_{H}^{2})\ .
\ee
Verify that the thermodynamic relation $\frac{\partial F}{\partial T}=-S$
agrees with the Bekenstein-Hawking formula for the black hole entropy.
Since this a first order phase transition you can also compute its
latent heat.
\end{exercise} 

In the last exercise, we saw that for a holographic CFT on a sphere
of radius $L$, the entropy is a discontinuous function of the temperature.
In fact, we found that for sufficiently high temperatures $T>\frac{d-1}{2\pi L}$,
the entropy was very large $S\sim C_{T}$, while for lower temperatures
the entropy was small because it did not scale with $C_{T}$. This
can be interpreted as deconfinement of the numerous degrees of freedom
measured by $C_{T}\gg1$ which do not contribute to the entropy below
the transition temperature $T_{c}=\frac{d-1}{2\pi L}$. How can this
bevavior be understood from the point of view of a large $N$ gauge
CFT?

\subsection{Applications\label{sub:Applications}}

The AdS/CFT correspondence (or the gauge/gravity duality more generally)
is a useful framework for thinking about strong coupling phenomena
in QFT. Besides the specific examples of strongly coupled CFTs that
can be studied in great detail using the gravitational dual description,
AdS/CFT provides a geometric reformulation of many effects in QFT.
Usually, we do not know the precise gravitational dual of a given
QFT of interest (like QCD) but it is still very useful to study gravitational
toy models that preserve the main features we are interested in. These
models enlarge our intuition because they are very different from
QFT models based on weakly interacting quasi-particles. There are
many examples of QFT observables that have a nice geometric interpretation
in the dual gravitational description. Perhaps the most striking one
is the computation of entanglement entropy as the area of a minimal
surface in the dual geometry \cite{hep-th/0603001}. Let us illustrate
this approach in the context of confinig gauge theories like pure
Yang-Mills theory.

Confinement means that the quark anti-quark potential between static
quarks grows linearly with the distance $L$ at large distances
\be
V(L)\approx\sigma L\ ,\qquad L\to\infty\ ,
\ee
where $\sigma$ is the tension of the flux tube or effective string.
This potential can be defined through the expectation value of a Wilson
loop (in the fundamental representation)
\be
W[C]={\rm Tr\ P}\exp\oint_{C}A_{\mu}dx^{\mu}\ ,
\ee
for a rectangular contour $C$ with sides $T\times L$,
\be
\left\langle W[C]\right\rangle \sim e^{-TV(L)}\ ,\qquad T\to\infty\:.
\ee
This is equivalent to the area law $\left\langle W[C]\right\rangle \sim e^{-\sigma Area[C]}$
for large contours. In the gauge/string duality there is a simple
geometric rule to compute expectation values of Wilson loops \cite{hep-th/9803002}.
One should evaluate the path integral 
\begin{equation}
\left\langle W[C]\right\rangle =\int_{\partial\Sigma=C}[d\Sigma]e^{-S_{s}[\Sigma]}\label{eq:stringpathintegral}
\end{equation}
summing over all surfaces $\Sigma$ in the dual geometry that end
at the contour $C$ at the boundary. The path integral is weighted
using the dual string world-sheet action. At large $N$, we expect
that the dominant contribution comes from surfaces $\Sigma$ with
disk topology. In specific examples, like SYM, this can be made very
precise. For example, at large 't Hooft coupling the world-sheet action
reduces to %
\footnote{In fact, the total area of $\Sigma$ is infinite but the divergence
comes from the region close to the boundary of AdS. This can be regulated
by cutting of AdS at $z=\epsilon$, and renormalized by subtracting
a divergent piece proportional to the length of the contour $C$.%
}
\begin{equation}
S_{s}[\Sigma]=\frac{1}{4\pi\ell_{s}^{2}}Area[\Sigma]\ .\label{eq:worldsheetAction}
\end{equation}
In this case, since the theory is conformal, there is no confinement
and the quark anti-quark potential is Coulomb like, 
\be
V(L)=\frac{a(N,\lambda)}{L}\ .
\ee

For most confining gauge theories (e.g. pure Yang-Mills theory) we
do not know neither the dual geometry nor the dual string world-sheet
action. However, we can get a nice qualitative picture if we assume
(\ref{eq:worldsheetAction}) and only change the background geometry.
The most general $(d+1)$-dimensional geometry that preserves $d$-dimensional
Poincaré invariance can be written as
\begin{equation}
ds^{2}=R^{2}\left[\frac{dz^{2}}{z^{2}}+A^{2}(z)dx^{\mu}dx_{\mu}\right]\ .\label{eq:confininggeometry}
\end{equation}
The profile of the function $A^{2}(z)$ encodes many properties of
the dual QFT. For a CFT, scale invariance fixes $A(z)\propto z^{-1}$.
For asymptotically free gauge theories, we still expect that $A(z)$
diverges for $z\to0$ however the function should be very different
for larger values of $z$. In particular, it should have a minimum
for some value $z=z_{\star}>0$. Let us see what this implies for
the expectation value of a large Wilson loop. The string path integral
(\ref{eq:stringpathintegral}) will be dominated by the surface $\Sigma$
with minimal area. For large contours $C$, this surface will sink
inside AdS until the value $z=z_{\star}$ that minimizes $A^{2}(z)$
and the worldsheet area will be given by 
\be
R^{2}A^{2}(z_{\star})Area[C]+O(Length[C])\ .
\ee
Therefore, we find a confining potential with flux tube tension
\be
\sigma=\frac{A^{2}(z_{\star})}{4\pi}\frac{R^{2}}{\ell_{s}^{2}}\ .
\ee

What happens if we put the QFT at finite temperature? In this case,
we can probe confinement by computing 
\be
\left\langle W(C_{x})\bar{W}(C_{x+L})\right\rangle _{\beta}=e^{-\beta F_{q\bar{q}}(\beta,L)}
\ee
where $C_{x}$ is the contour around the Euclidean time circle at the
spatial position $x$ (Polyakov loop). $F_{q\bar{q}}(\beta,L)$ denotes
the free energy of a static quark anti-quark pair at distance $L$
and temperature $1/\beta$. If $F_{q\bar{q}}(\beta,L)\to\infty$ as
we separate the pair, then we are in the confined phase. On the other
hand, if $F_{q\bar{q}}(\beta,L)$ remains finite when $L\to\infty$,
we are in the deconfined phase. Let us see how this works in the holographic
dual. For low temperatures, the dual geometry is simply given by (\ref{eq:confininggeometry})
with Euclidean time identified with period $\beta.$ Therefore, the
bulk minimal surface that ends on $C_{x}$ and $C_{x+L}$ will have
a cylindrical topology and its area will scale linearly with $L$
at large $L$. In fact, we find $F_{q\bar{q}}(\beta,L)\approx\sigma L$
like in the vacuum. On the other hand, for high enough temperature
we expect the bulk path integral to be dominated by a black hole geometry
(see exercise \ref{ex:HPtransition} about Hawking-Page phase transition). The metric can
then be written as 
\begin{equation}
ds^{2}=R^{2}\left[\frac{dz^{2}}{z^{2}f(z)}+f(z)d\tau^{2}+g(z)dx^{i}dx_{i}\right]\ ,\label{eq:confininggeometry-1}
\end{equation}
where $f(z)$ vanishes for some value $z=z_{H}$. This means that
the Euclidean time circle is contractible in the bulk. Therefore,
for large $L$, the minimal surface has two disconnected pieces with
disk topology ending on $C_{x}$ and $C_{x+L}$ whose area remains
finite when $L\to\infty$. This means deconfinement
\be
\lim_{L\to\infty}\left\langle W(C_{x})\bar{W}(C_{x+L})\right\rangle _{\beta}=\left\langle W(C_{x})\right\rangle _{\beta}^{2}=e^{-2\beta F_{q}(\beta)}>0\ .
\ee

Another feature of a confining gauge theory is a mass gap and a discrete
spectrum of mesons and glueballs. To compute this spectrum using the
bulk dual one should study fluctuations around the vacuum geometry
(\ref{eq:confininggeometry}). Consider for simplicity, a scalar field
obeying $\nabla^{2}\phi=m^{2}\phi$. Since we are interested in finding
the spectrum of the operator $P_{\mu}P^{\mu}$ we look for solutions
of the form $\phi=e^{ik\cdot x}\psi(z)$, which leads to 
\begin{equation}
\frac{z}{A^{d}(z)}\partial_{z}\left(zA^{d}(z)\partial_{z}\psi\right)-\frac{k^{2}}{A^{2}(z)}\psi=m^{2}R^{2}\psi\ .\label{eq:RadialeqConfgeo}
\end{equation}
The main idea is that this equation will only have solutions that
obey the boundary conditions $\psi(0)=\psi(\infty)=0$ for special
discrete values of $k^{2}$. In other words, we obtain a discrete
mass spectrum as expected for a confining gauge theory.
\begin{exercise} 
Consider the simplest holographic model of a confining
gauge theory: the hard wall model. This is just a slice of AdS, i.e.
we take $A(z)=1/z$ and cutoff space at $z=z_{\star}$. Show that
(\ref{eq:RadialeqConfgeo}) reduces to the Bessel equation
\be
\left[z^{2}\partial_{z}^{2}+z\partial_{z}-\alpha^{2}-k^{2}z^{2}\right]h(z)=0\ ,
\ee
where $\alpha^{2}=m^{2}R^{2}+d^{2}/4$ and $h(z)=z^{-\frac{d}{2}}\psi(z)$.
Finally, show that the boundary conditions $h(0)=h(z_{\star})=0$,
lead to the quantization
\be
h_{n}(z)=J_{\alpha}\left(\frac{z}{z_{\star}}u_{\alpha,n}\right)\ ,\qquad m_{n}^{2}=-k^{2}=\frac{u_{\alpha,n}^{2}}{z_{\star}^{2}}\ ,\qquad n=1,2,\dots
\ee
where $u_{\alpha,n}$ is the $n$th zero of the Bessel function $J_{\alpha}$. 
\end{exercise}

It is instructive to compare the lightest glueball mass $m_{1}$ with
the flux tube tension $\sigma=\frac{1}{4\pi z_{\star}^{2}}\frac{R^{2}}{\ell_{s}^{2}}$
in the hard wall model. We find that $\frac{\sigma}{m_{1}^{2}}\sim\frac{R^{2}}{\ell_{s}^{2}}$.
The fact that this ratio is of order 1 in pure Yang-Mills theory is
another indication that its holographic dual must be very stringy
(curvature radius of the same order of the string length). 

Above the deconfinement temperature, the system is described by a
plasma of deconfined partons (quarks and gluons in QCD). The gauge/gravity
duality is also very useful to describe this strongly coupled plasma.
The idea is that the hydrodynamic behavior of the plasma is dual to
the long wavelength fluctuations of the black hole horizon. This map
can be made very precise and has led to significant developments in
the theory of relativistic hydrodynamics. One important feature of
the gravitational description is that dissipation is built in because
black hole horizons naturally relax to equilibrium. A famous result
from this line of work was the discovery of a universal ratio of shear
viscosity $\eta$ to entropy density $s$. Any CFT dual to Einstein
gravity in AdS has $\frac{\eta}{s}=\frac{1}{4\pi}$. This is a rather
small number (water at room temperature has $\frac{\eta}{s}\sim30$)
but remarkably it is of the same order of magnitude of that observed
in the quark-gluon plasma produced in heavy ion collisions \cite{arXiv:0706.1522}.

There are also many interesting applications of the gauge/gravity
duality to Condensed Matter physics \cite{arXiv:0903.3246, arXiv:0909.0518}.
There are many materials that are not well described by weakly coupled quasi-particles. In this case, it is useful to have alternative models based on gravitational theories in AdS that  share the same qualitative features.
This can give geometric intuition about the system in question.

The study of holographic models is also very useful for the discovery
of general properties of CFT (and QFT more generally). If one observes
that a given property holds both in weakly coupled and in holographic
CFTs, it is natural to conjecture that such property holds in all
CFTs. This reasoning has led to the discovery (and sometimes proof)
of several important facts about CFTs, like the generalization of
Zamolodchikov's c-theorem to $d>2$ (known as F-theorem in $d=3$
and a-theorem in $d=4$) \cite{Zamolodchikov:1986gt, arXiv:1011.5819,  arXiv:1107.3987 , arXiv:1202.5650}
or the existence of universal bounds on the three-point function of
the stress tensor and its relation to the idea of energy correlators
\cite{arXiv:0803.1467, arXiv:1603.03771, arXiv:1605.08072}.

Onother example along this line is the existence of ``double-trace''
operators with large spin in any CFT. The precise statement is that
in the OPE of two operators $\mathcal{O}_{1}$ and $\mathcal{O}_{2}$
there is an infinite number of operators $\mathcal{O}_{n,l}$ of spin
$l\gg1$ and scaling dimension

\begin{equation}
\Delta_{n,l}\approx\Delta_{1}+\Delta_{2}+2n+l+\frac{\gamma_{n}}{l^{\tau_{min}}}\label{eq:LargeSpinDoubleTraces}
\end{equation}
where $\tau_{min}$ is the minimal twist (dimension minus spin) of
all the operators that appear in both OPEs $\mathcal{O}_{1}\times\mathcal{O}_{1}$
and $\mathcal{O}_{2}\times\mathcal{O}_{2}$. In a generic CFT, this
will be the stress tensor with $\tau_{min}=d-2$ and one can derive
explicit formulas for $\gamma_{n}$ \cite{arXiv:1212.3616, arXiv:1212.4103, arXiv:1502.01437, arXiv:1504.00772}.
This statement has been proven using the conformal bootstrap equations
but its physical meaning is more intuitive in the dual AdS language.
Consider two particle primary states in AdS. Without interactions
the energy of such states is given by $\Delta_{1}+\Delta_{2}+2n+l$
where $n=0,1,2,\dots$ is a radial quantum number and $l$ is the
spin. Turning on interactions will change the energies of these two-particle
states. However, the states with large spin and fixed $n$ correspond
to two particles orbitating each other at large distances and therefore
they will suffer a small energy shift due to the gravitational long
range force. At large spin, all other interactions (corresponding
to operators with higher twist) give subdominant contributions to
this energy shift. In other words, the general result (\ref{eq:LargeSpinDoubleTraces})
is the CFT reflection of the simple fact that interactions decay with
distance in the dual AdS picture.

\section{Mellin amplitudes \label{sec:Mellin}}

Correlation functions of local operators in CFT are rather complicated functions of the cross-ratios. Since these are crucial observables in AdS/CFT it is useful to find simpler representations. This is the motivation to study
Mellin amplitudes.
They were introduced by G. Mack in 2009 \cite{Mack:2009mi, Mack:2009gy} following earlier work \cite{Symanzik:1972wj, Dobrev:1975ru}.
Mellin amplitudes share many of the properties of scattering amplitudes of dual resonance models. In particular, they are crossing symmetric and have a simple analytic structure (related to the OPE). As we shall see, in the case of holographic CFTs, we can take this analogy  further and obtain bulk flat space scattering amplitudes as a limit of the dual CFT Mellin amplitudes.
Independently of AdS/CFT applications, Mellin amplitudes can be useful to describe CFTs in general.

\subsection{Definition}

Consider the $n$-point function of scalar primary operators \footnote{We shall use the notation $M(\gamma_{ij})$ to denote a function $M(\gamma_{12},\gamma_{13},\dots)$ of all Mellin variables.}
\be
\left\langle \mathcal{O}_1(P_1) \dots \mathcal{O}_n(P_n) \right\rangle =
\int [d\gamma] M(\gamma_{ij}) \prod_{1\le i < j \le n} \frac{\Gamma(\gamma_{ij})}{(-2P_i\cdot P_j)^{\gamma_{ij}}}
\label{eq:Mellindef}
\ee
Conformal invariance requires weight $-\Delta_i$ in each $P_i$. This leads to constraints in the Mellin variables which can be conveniently written as
\be
\sum_{j=1}^n \gamma_{ij} =0\ ,\qquad \gamma_{ij}=\gamma_{ji}\ ,\qquad \gamma_{ii}=-\Delta_i\ .
\label{gammaconstraints}
\ee
Notice that for $n=2$ and $n=3$ the Mellin variables are entirely fixed by these constraints. In these cases, there is no integral to do and the Mellin representation just gives the known form of the conformal two and three point function.
The integration measure $[d\gamma]$ is over the $n(n-3)/2$ independent Mellin variables (including a factor of $\frac{1}{2\pi i}$ for each variable) and the integration contours run parallel to the imaginary axis. The precise contour in the complex plane is dictated by the requirement that it should pass to the right/left of the semi-infinite sequences of poles of the integrand that run to the left/right. This will become clear in the following example.

\begin{figure}
\begin{centering}
\includegraphics[clip,width=0.6\textwidth]{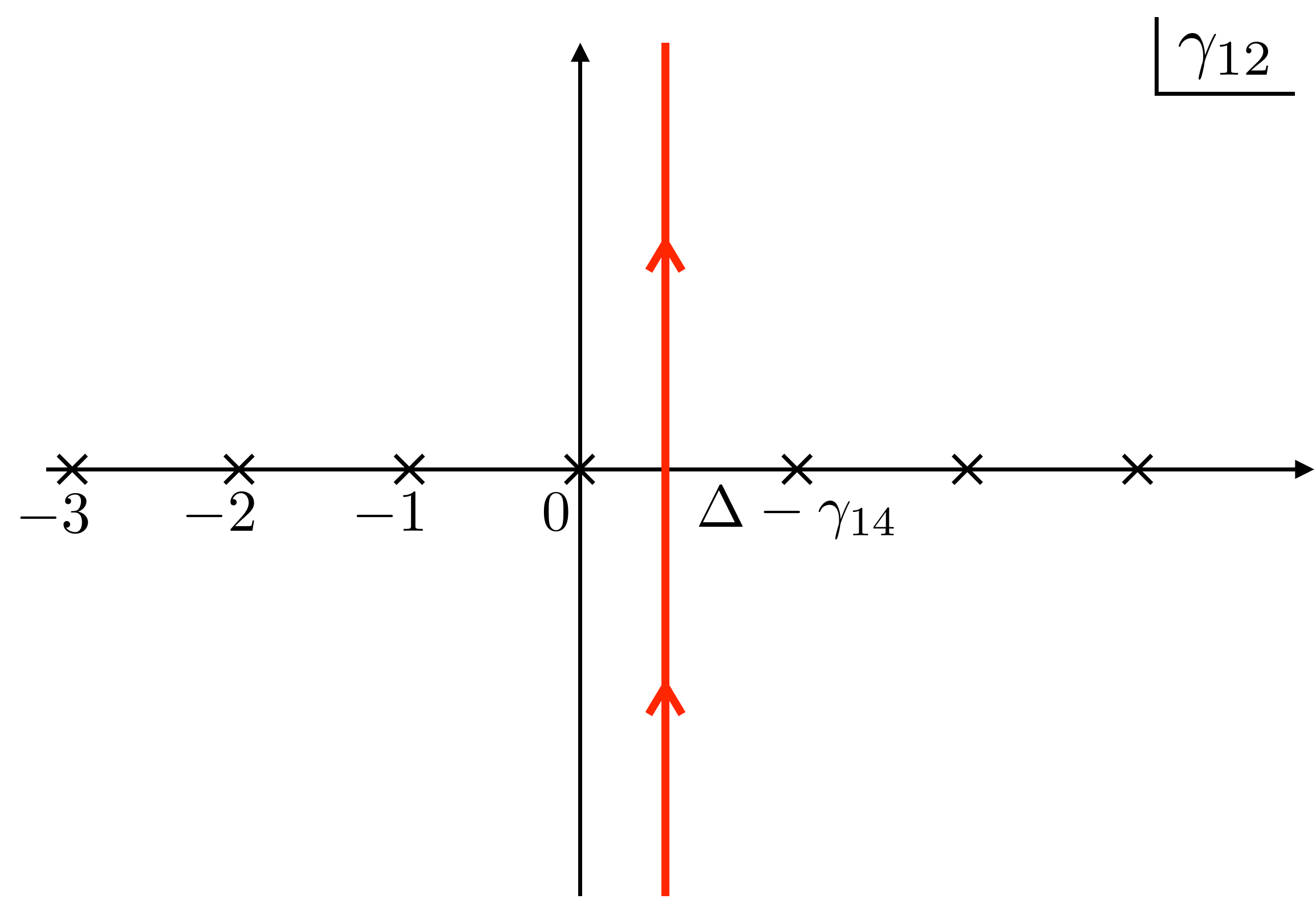}
\caption{Integration contour for the Mellin variable $\gamma_{12}$. The crosses represent  (double) poles of the $\Gamma$-functions given by \eqref{eq:gamma12poles1} and  \eqref{eq:gamma12poles2}. In general, the Mellin amplitude  has several semi-infinite sequence of poles. Each sequence should stay entirely on one side of the contour.
\label{fig:gamma12plane}}
\end{centering}
\end{figure}

Consider the case of a four-point function of a scalar operator of dimension $\Delta$. In this case, there are two independent Mellin variables which we can choose to be $\gamma_{12}$ and $\gamma_{14}$. This leads to
\be
\left\langle \mathcal{O}(P_1) \dots \mathcal{O}(P_4) \right\rangle =
\frac{1}{(P_{13}P_{24})^\Delta}
\int_{-i\infty}^{i\infty} \frac{d\gamma_{12}\gamma_{14}}{(2\pi i)^2}   \hat{M}(\gamma_{12}, \gamma_{14}) u^{-\gamma_{12}}  v^{-\gamma_{14}} \,,
\ee
where $u$ and $v$ are the cross ratios (\ref{crossratios}) and 
\be
\hat{M}(\gamma_{12}, \gamma_{14})=M(\gamma_{12}, \gamma_{14})
 \Gamma^2(\gamma_{12})
 \Gamma^2(\gamma_{14})
  \Gamma^2(\Delta-\gamma_{12}-\gamma_{14})\,.
\ee
Consider the first the complex plane of $\gamma_{12}$ depicted in figure \ref{fig:gamma12plane}.
The $\Gamma$-functions give rise to semi-infinite sequences of (double) poles at
\begin{align}
\gamma_{12}&=0, -1, -2, \dots  \label{eq:gamma12poles1}\\
\gamma_{12}&=\Delta-\gamma_{14}, \Delta-\gamma_{14}+1, \Delta-\gamma_{14}+2, \dots 
\label{eq:gamma12poles2}
\end{align}
As we shall see in the next section, the Mellin amplitude $M(\gamma_{ij})$ also has the same type of semi-infinite sequences of poles. The integration contour should pass in the middle of these sequences of poles as shown in figure \ref{fig:gamma12plane}.
Invariance of the four-point function under permutation of the insertion points $P_i$, leads to crossing symmetry of the Mellin amplitude
\be
M(\gamma_{12}, \gamma_{13}, \gamma_{14}) =
M(\gamma_{13}, \gamma_{12}, \gamma_{14}) =
M(\gamma_{14}, \gamma_{13}, \gamma_{12}) \ ,
\ee
where we used 3 variables obeying a single constraint $\gamma_{12}+ \gamma_{13}+ \gamma_{14}=\Delta$.
This is reminiscent of crossing symmetry of scattering amplitudes written in terms of Mandelstam invariants.

It is convenient to introduce fictitious momenta $p_i$ such that $\gamma_{ij}=p_i\cdot p_j$. 
Imposing momentum conservation $\sum_{i=1}^n p_i=0$ and the on-shell condition $p_i^2=-\Delta_i$ automatically leads to the constraints (\ref{gammaconstraints}).
These fictitious momenta are a convenient trick but we do not know how to define them directly.
In all formulas, we will only use their inner products $\gamma_{ij}=p_i\cdot p_j$. 
In particular, it is not clear in what vector space do the momenta $p_i$ live. 
\footnote{The flat space limit of AdS discussed in section \ref{sec:FSL}, suggests a $d+1$ dimensional space but this is unclear before the limit.}

Let us be more precise about the number of independent cross ratios.
The correct formula is
\begin{align}
&\frac{n(n-3)}{2}\ ,\qquad &n\le d+2 \\
& n d  - \frac{(d+1)(d+2)}{2}\ ,\qquad  &n\ge d+2 
\end{align}
In fact, for $n> d+2$ one can write identities like
\be
\det_{i,j} P_i \cdot P_j = 0
\ee
using $d+3$ embedding space vectors.
Notice that this makes the Mellin representation non-unique.
We can shift the Mellin amplitude by the Mellin transform of 
\be
F(P_1, \dots, P_n)  \det_{i,j} P_i \cdot P_j =0
\ee
where $F$ is any scalar function with the appropriate homogeneity properties.
This non-uniqueness of the Mellin amplitude is analogous to the non-uniqueness of the $n$-particle scattering amplitudes (as functions of the invariants $k_i\cdot k_j$) in $(d+1)$-dimensional spacetime if $n>d+2$.

\subsection{OPE $\Rightarrow$ Factorization} 
Consider the OPE
\be
\mathcal{O}_1(x_1) \mathcal{O}_1(x_2) =
\sum_k C_{12k} \left(x_{12}^2 \right)^{\frac{\Delta_k-\Delta_1-\Delta_2}{2}}
\left[
\mathcal{O}_k(x_2)+
c\, x_{12}^2 \partial^2 \mathcal{O}_k(x_2)
+\dots
\right] 
\ee
where the sum is over primary operators $\mathcal{O}_k$ and, for simplicity, we wrote the contribution of a scalar operator. The term proportional to the constant $c$ is a descendant and is fixed by conformal symmetry like all the other terms represented by $\dots$.
Let us compare this with the Mellin representation. 
When $x_{12}^2  \to 0$, it is convenient to integrate over $\gamma_{12}$ closing the contour to the left in the $\gamma_{12}$-complex plane. This gives
\be
\langle 
\mathcal{O}_1(x_1) \mathcal{O}_1(x_2) \dots \rangle
= \sum_{\bar{\gamma}_{12}} \left( x_{12}^2  \right)^{-\bar{\gamma}_{12}}
\int [d\gamma]' {\rm Res}_{\bar{\gamma}_{12}} \hat{M}(\gamma_{ij})
\prod' \left(x_{ij}^2 \right)^{-\gamma_{ij}}
\ee
where $[d\gamma]'$ and $\prod'$ stand for the integration measure and product excluding $ij=12$.
Comparing the two expressions we conclude that $\hat{M}$ must have poles at 
\be
\gamma_{12}=\frac{\Delta_1+\Delta_2-\Delta_k-2m}{2}\ ,\qquad  \qquad
m=0,1,2,\dots
\ee
where the poles with $m>0$ correspond to descendant contributions.
If the CFT has a discrete spectrum of scaling dimensions then its Mellin amplitudes are analytic functions with single poles as its only singularities (meromorphic functions).
It is also clear that the residues of these poles will be proportional to the product of the OPE coefficient $C_{12k}$ and the Mellin amplitude of the lower point correlator $\langle \mathcal{O}_k \dots \rangle$.
The precise formulas are derived in \cite{Mack:2009mi, Goncalves:2014rfa}. Here we shall just list the main results without derivation.

\subsubsection{Four-point function \label{sec:MellinOPE4pt}}

In the case of the four-point function it is convenient to write the Mellin amplitude in terms of `Mandelstam invariants'
\begin{align}
s&=-(p_1+p_2)^2=\Delta_1+\Delta_2-2\gamma_{12}\\
t&=-(p_1+p_3)^2=\Delta_1+\Delta_3-2\gamma_{13}
\end{align}
Then, the poles and residues of the Mellin amplitude take the following form \cite{Mack:2009mi}
\be
M(s,t) \approx C_{12k} C_{34k}  \frac{
Q_{l_k,m}(t)}{s-\Delta_k+l_k-2m}\ ,\qquad m=0,1,2,\dots
\label{eq:polesMellin4pt}
\ee
where $Q_{l,m}(t)$ is a kinematical polynomial of degree $l$ in the variable $t$.

This strengthens  the analogy with scattering amplitudes. Each operator of spin $l$ in the OPE $\mathcal{O}_1\times
\mathcal{O}_2$ gives rise to poles in the Mellin amplitude very similar to the poles in the scattering amplitude associated to the exchange of a 
particle of the same spin.

\subsubsection{Planar correlators}

Notice that the polynomial behaviour of the residues requires the inclusion of the $\Gamma$-functions in the definition (\ref{eq:Mellindef}) of Mellin amplitudes.
On the other hand, the $\Gamma$-functions themselves have poles at fixed positions. For example, $\Gamma(\gamma_{12})$ gives rise to poles at $s=\Delta_1+\Delta_2+2m$ with $m=0,1,2,\dots$.
In a generic CFT, there are no operators with these scaling dimensions and therefore the Mellin amplitude must have zeros at these values to cancel these unwanted OPE contributions.
However, in correlation functions of single-trace operators in large $N$ CFTs we expect precisely this type of contributions. At the planar level, the $\Gamma$-functions account for all multi-trace OPE contributions and the Mellin amplitude only has poles associated to  single-trace  operators.

\subsubsection{$n$-point function}

Considering the OPE of $k$ scalar operators, one can derive more general factorization formulas \cite{Goncalves:2014rfa}. For example, for each primary operator $\Op$ of dimension $\D$ and spin $l$ that appears in the OPEs $\Op_1 \times \dots \times \Op_k$ and $\Op_{k+1} \times \dots \times \Op_n$, we obtain the following sequence of poles in the $n$-point Mellin amplitude,
\be
M_n \approx \frac{Q_m}{\gamma_{LR} -\Delta+l-2m}\,,\qquad
m=0,1,2,\dots
\ee
where
\be
\gamma_{LR}=-\left(\sum_{i=1}^k p_i \right)^2=\sum_{i=1}^k \sum_{j>k}^n \gamma_{ij}\,.
\ee
In general, the residue can be written in terms of lower point Mellin amplitudes. For example, if $l=0$ the residue factorizes
\be
Q_0 = -2\Gamma(\Delta) M^L_{k+1} \, M^R_{n-k+1}\,,
\label{eq:facMellinm0}
\ee
with $M^L_{k+1}$ the Mellin amplitude of $\langle \mathcal{O}_1\dots \mathcal{O}_k \mathcal{O}\rangle$ and $M^R_{n-k+1}$   the Mellin amplitude of $\langle  \mathcal{O} \mathcal{O}_{k+1}\dots \mathcal{O}_n\rangle$ .
The satellite poles also factorize but give rise to more complicated formulae
\be
Q_m = \frac{-2\Gamma(\Delta)m!}{\left(\Delta-\frac{d}{2}+1\right)_m} L_m \, R_m\,,
\label{eq:facMellinm}
\ee
with
\be
L_m=\sum_{ n_{ab}\ge 0 \atop \sum n_{ab}=m} M^L(\gamma_{ab}+n_{ab}) 
\prod_{1\le a<b \le k} \frac{(\gamma_{ab})_{n_{ab}}}{n_{ab}!}
\ee
and similarly for $R_m$.

There also factorization formulas for the residues associated with operators with non-zero spin \cite{Goncalves:2014rfa}. 
However, the general case including external operators with spin has not been worked out.

\subsection{Holographic CFTs}

As discussed in section \ref{sec:QGinAdS}, holographic CFTs have two special properties: large $N$ factorization and a small number of low dimension single-trace operators. Therefore, one should expect that the corresponding Mellin amplitudes are particularly simple, at least at the planar level. We shall now confirm this expectation with a few simple examples.

\begin{figure}
\begin{centering}
\includegraphics[clip,width=0.4\textwidth]{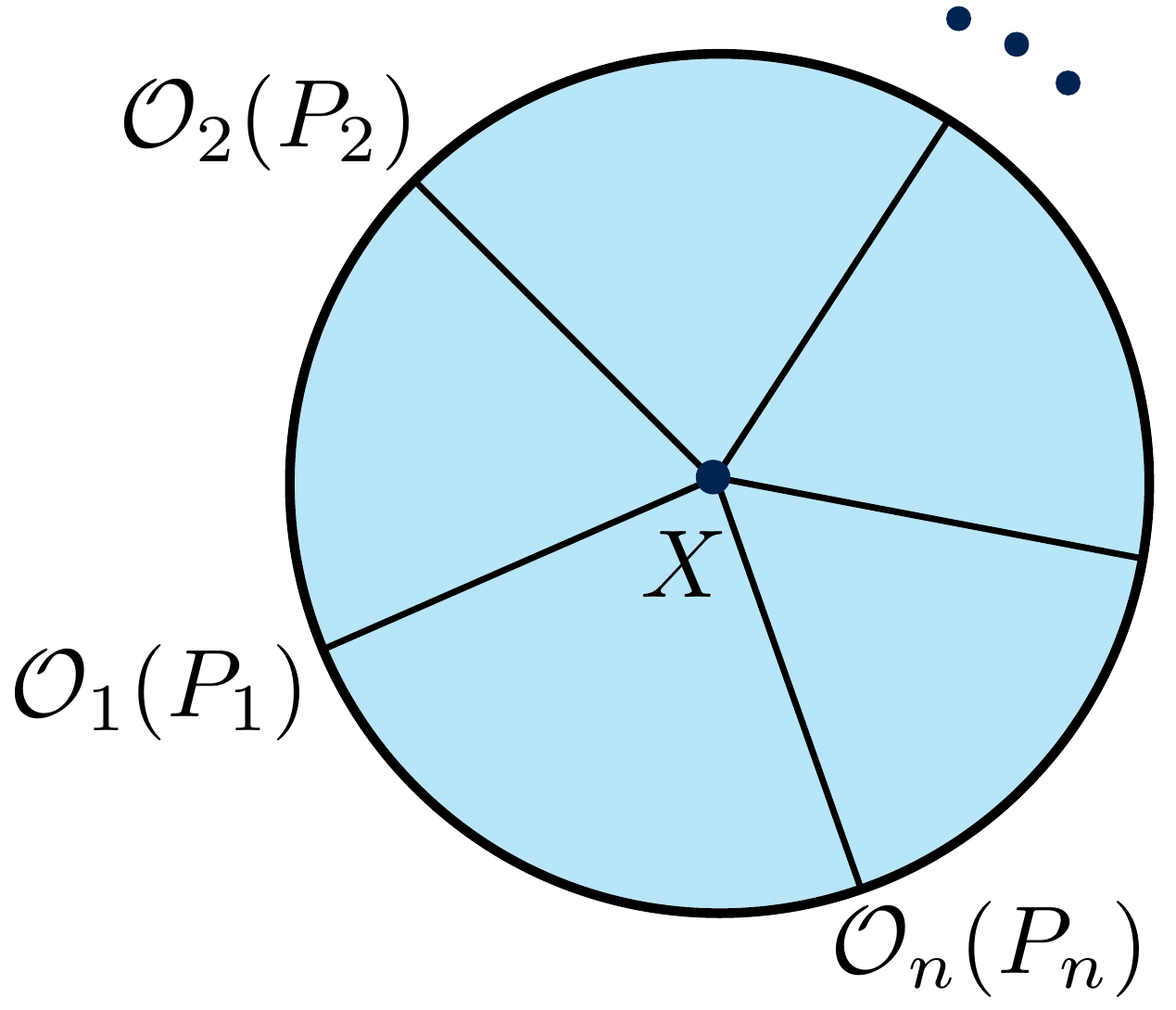}
\caption{Witten diagram for a $n$-point contact interaction in AdS. 
The interior of the disk represents the bulk of AdS and the circumference represents its conformal boundary. The lines connecting the boundary points $P_i$ to the bulk point $X$ represent bulk to boundary propagators. 
\label{fig:AdScontact}}
\end{centering}
\end{figure}

\subsubsection{Witten diagrams}

Consider the contact Witten diagram of figure \ref{fig:AdScontact}. It corresponds to an interaction vertex $\lambda \phi_1 \dots \phi_n$ in the bulk lagrangian and it contributes \footnote{We are using CFT operators $\Op_i$ normalized to have unit two point function.}
\be
\langle \Op_1(P_1) \dots \Op_n(P_n) \rangle = \lambda \int_{AdS} dX \prod_{i=1}^n \frac{ \sqrt{\mathcal{C}_{\D_i}}}{
(-2P_i \cdot X)^{\D_i}} \label{AdScontact}
\ee
to the dual CFT correlation function.
One can show that this corresponds to a constant Mellin amplitude,
\be
M=\lambda\, \frac{1}{2}\pi^{\frac{d}{2}}\Gamma\left(\frac{\sum \D_i -d}{2} \right)
\prod_{i=1}^n \frac{ \sqrt{\mathcal{C}_{\D_i}}}{\Gamma(\D_i)}\ .
\ee
\begin{exercise}
\label{ex:Mellincontact}
Check the last statement. Start by using the integral representation of the bulk to boundary propagators and performing the integral over AdS using Poincare coordinates as explained in exercise \ref{ex:3pt}. This  turns (\ref{AdScontact}) into
\be
\lambda\, \pi^{\frac{d}{2}}\Gamma\left(\frac{\sum \D_i -d}{2} \right) 
\int_0^\infty  e^{-\sum_{i<j} s_i s_j P_{ij}}
\prod_{i=1}^n \frac{ \sqrt{\mathcal{C}_{\D_i}}}{\Gamma(\D_i)}
s_i^{\D_i-1} ds_i \ .
\ee
Next, use the Mellin representation ($c>0$)
\be
e^{-s_i s_j P_{ij}} =\int_{c-i\infty}^{c+i\infty} \frac{d\gamma_{ij}}{2\pi i} \Gamma(\gamma_{ij})  (s_i s_j P_{ij})^{-\gamma_{ij}}
\ee
for $n(n-3)/2$ exponential factors. A good choice is to keep $n$ factors, corresponding to the exponential
\be
e^{-s_1\sum_{i=2}^n  s_{i} P_{1i}-s_2s_3 P_{23}}\,.
\ee
The integrals over $s_4,\dots,s_n$ can be easily done in terms of $\Gamma$-functions. Finally, do the integrals over $s_1,s_2,s_3$ using the same type of change of variables as in exercise \ref{ex:3pt}. 
\end{exercise}

This result can be easily generalized to interaction vertices with derivatives. For example, the vertex $\lambda (\nabla_\alpha \phi_1 \nabla^\alpha \phi_2) \phi_3 \dots \phi_n$ gives rise to 
\begin{align}
&\langle \Op_1(P_1) \dots \Op_n(P_n) \rangle =
\lambda \int_{AdS} dX 
\prod_{i=3}^n \frac{ \sqrt{\mathcal{C}_{\D_i}}}{
(-2P_i \cdot X)^{\D_i}} \times\label{AdScontactder} \\
&\qquad\qquad \times
(\eta^{AB}+X^AX^B)
\frac{\partial}{\partial X^A} \frac{ \sqrt{\mathcal{C}_{\D_1}}}{
(-2P_1 \cdot X)^{\D_1}}
\frac{\partial}{\partial X^B}
\frac{ \sqrt{\mathcal{C}_{\D_2}}}{
(-2P_2 \cdot X)^{\D_2}}\,.
\nonumber
\end{align}
Here we have used the fact that covariant derivatives in AdS can be computed as partial derivatives in the embedding space projected to the AdS sub-manifold.\footnote{See appendix F.1 of \cite{Costa:2016hju} for a derivation of this statement in the analogous case of a sphere embedded in Euclidean space.}
This gives
\be
\lambda \D_1 \D_2 \left( -2P_{12}\, D_{\Delta_1+1, \Delta_2+1, \Delta_3, \dots,\D_n} +
  D_{\Delta_1, \Delta_2, \Delta_3, \dots, \D_n}\right)
  \prod_{i=1}^n \sqrt{\mathcal{C}_{\D_i}}
\ee
where we introduced the D-function \cite{hep-th/9903196}
\be
D_{\Delta_1,  \dots, \D_n} \equiv  \int_{AdS} dX 
\prod_{i=1}^n \frac{1}{
(-2P_i \cdot X)^{\D_i}} \,.
\ee
More generally, it is clear that the contact Witten diagram associated with a generic vertex 
$\lambda  \nabla  \dots \nabla  \phi_1  \nabla  \dots \nabla \phi_2    \dots  \nabla  \dots \nabla\phi_n$ with all derivatives contracted among different fields, gives rise to a linear combination of terms of the form
\be
D_{\Delta_1+\Lambda_1,  \dots, \D_n+\Lambda_n} \prod_{i<j}^n P_{ij}^{\lambda_{ij}}
\label{shiftedDfunctions}
\ee
where $\lambda_{ij}$ are non-negative integers and $\Lambda_i=\sum_{j\neq i} \lambda_{ij}$. 
As we will see in the next exercise, the Mellin amplitude of (\ref{shiftedDfunctions}) is a polynomial in the Mellin variables. Therefore, the Mellin amplitude associated to contact Witten diagrams is polynomial. The absence of poles in the Mellin amplitude means that the conformal block decomposition of the contact diagram only contains multi-trace operators, in agreement with previous results \cite{Liu:1998th, D'Hoker:1998mz}.

\begin{exercise} \label{ex:generalcontact}
If the vertex $\lambda  \nabla  \dots \nabla  \phi_1  \nabla  \dots \nabla \phi_2    \dots  \nabla  \dots \nabla\phi_n$ has $2N=2\sum_{i<j} \alpha_{ij} $ derivatives with $\alpha_{ij}$ contractions of derivatives acting on $\phi_i$ and $\phi_j$, show that the contact Witten diagram is given by
\be
\lambda \left(\prod_{i=1}^n \sqrt{\mathcal{C}_{\D_i}} \right)
D_{\Delta_1+\Lambda_1,  \dots, \D_n+\Lambda_n} \prod_{i<j}^n (-2P_{ij})^{\alpha_{ij}} +\dots
\label{eq:contactwithder}
\ee
where $\Lambda_i=\sum_{j\neq i} \alpha_{ij}$ and the $\dots$ represent similar terms with less $P_{ij}$ factors. Hint: use the trick of writing covariant derivatives in AdS as partial derivatives in the embedding space projected to the AdS sub-manifold. %Then, the term with maximal factors of $P_{ij}$ follows from using always the first term in the projector $\eta^{AB}+X^A X^B$.

The Mellin representation of the $D$-functions is very simple. As we saw in exercise \ref{ex:Mellincontact}, the Mellin amplitude associated to $D_{\Delta_1,  \dots, \D_n}$  is simply
\be
\frac{\pi^\frac{d}{2} \Gamma\left( \frac{\sum \D_i -d}{2}\right)}
{2\prod_{i=1}^n \Gamma(\D_i)}\ .
\ee
Show that the Mellin amplitude associated to the correlation function (\ref{eq:contactwithder}) is given by the polynomial
\be
\lambda \left(\prod_{i=1}^n \sqrt{\mathcal{C}_{\D_i}} \right)
\frac{\pi^\frac{d}{2} \Gamma\left( \frac{\sum \D_i+2N -d}{2}\right)}
{2\prod_{i=1}^n \Gamma(\D_i+\Lambda_i)}
 \prod_{i<j}^n (-2\g_{ij})^{\alpha_{ij}} +\dots
 \label{Mellingeneralcontact}
\ee
where  the $\dots$ represent terms of lower degree in $\g_{ij}$. Hint: this follows easily from shifting the integration variables in the Mellin representation (\ref{eq:Mellindef}).
\end{exercise}

\begin{figure}
\begin{centering}
\includegraphics[clip,width=0.5\textwidth]{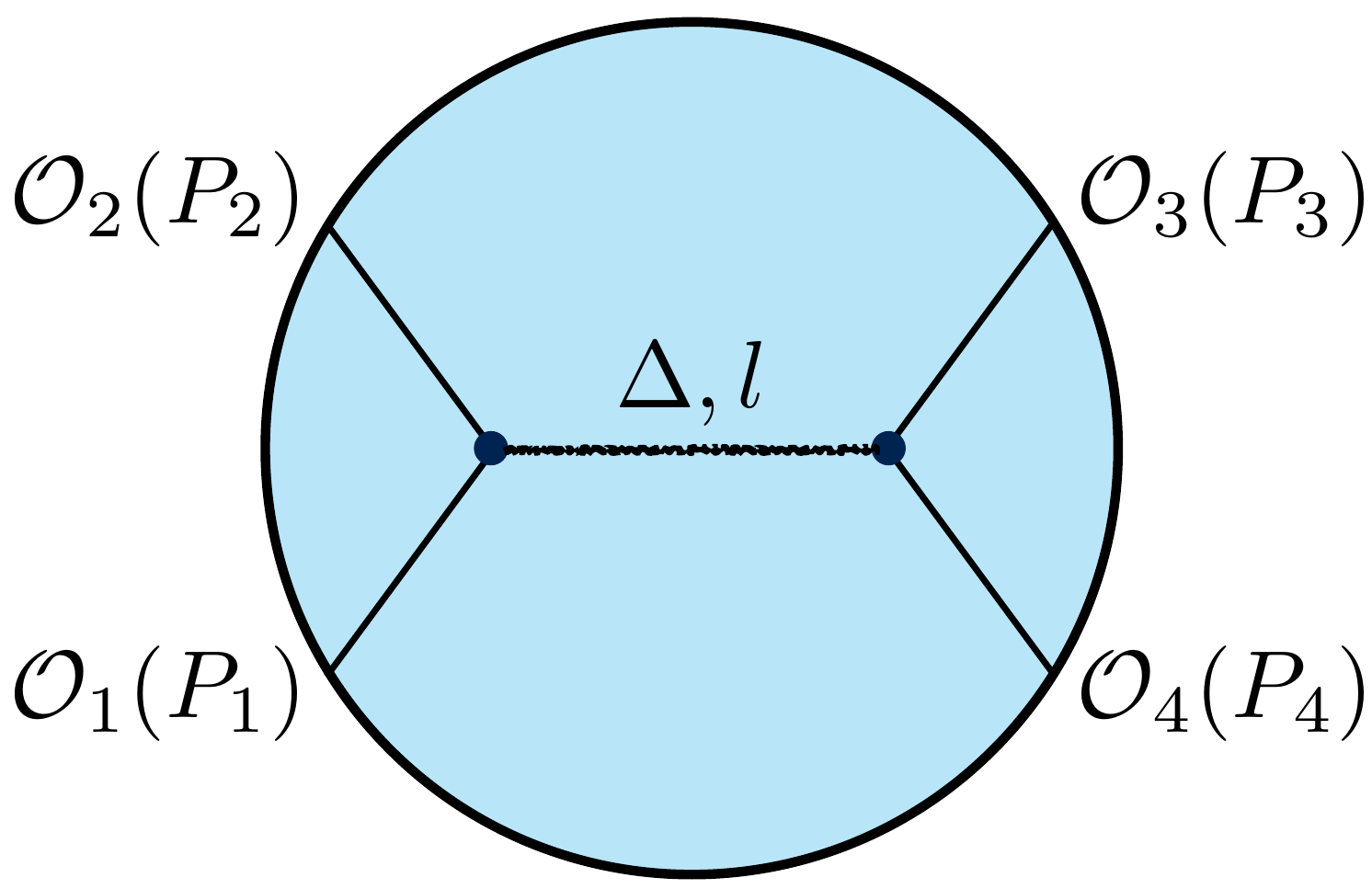}
\caption{Witten diagram describing the exchange of a bulk field dual to an operator of dimension $\Delta$ and spin $l$. 
\label{fig:AdSexchange}}
\end{centering}
\end{figure}

Consider now the Witten diagram shown in figure \ref{fig:AdSexchange} describing the exchange of a bulk field dual to a single-trace boundary operator $\Op$ of dimension $\Delta$ and spin $l$.
The conformal block decomposition of this diagram in the (12)(34) channel contains the single-trace operator  $\Op$  plus  double-trace operators schematically of the form $\Op_1 (\partial^{2})^n\partial_{\mu_1\dots \mu_j}\Op_2$ and $\Op_3 (\partial^{2})^n\partial_{\mu_1\dots \mu_j}\Op_4$. Moreover, the OPE in the crossed channels only contains double-trace operators.
This means that the Mellin amplitude is of the form
\be
M=C_{12\Op}C_{34\Op} \sum_{m=0}^\infty \frac{Q_{l,m}(t)}{s-\D+l-2m} + R(s,t)
\ee
where the OPE coefficients $C_{12\Op}$ and $C_{34\Op}$ are proportional to the bulk cubic couplings and $R(s,t)$ is an analytic function. 
The residues are proportional to degree $l$ Mack polynomials $Q_{l,m}(t)$ which are entirely fixed by conformal symmetry as we saw in \ref{sec:MellinOPE4pt}.
If we choose minimal coupling between the spin $l$ bulk field and the external scalars, then $R(s,t)$ is a polynomial of degree $\le l-1$.
This is particularly simple in the case of a scalar exchange ($l=0$). Then the residues are independent of $t$ and $R=0$ \cite{arXiv:1011.1485}.
Notice that this simple looking Mellin amplitude gives rise to a rather involved function of the cross-ratios in position space. This example  illustrates clearly the advantage of using the Mellin reprsentation to describe Witten diagrams.

The Mellin amplitude of a general tree-level scalar Witten diagrams was determined in \cite{arXiv:1107.1499, Paulos:2011ie, arXiv:1111.6972, arXiv:1112.0305}. The final result can be summarized in the following Feynman rules:
\begin{itemize}
\item Associate a momentum $p_j$ to every line (propagator) in the Witten diagram. External lines have incoming momentum $p_i$ satisfying $-p_i^2=\D_i$. Momentum is conserved at every vertex of the diagram.
\item Assign an integer $m_j$ to every line. External lines have $m_i=0$.
\item Every internal line (bulk-to-bulk propagator) contributes a factor\footnote{
The propagator numerator is given by
\be
S_{m}^{\D}=
\frac{\Gamma\left(\Delta-\frac{d}{2}+1+m\right)}{2(m!) \Gamma^2\left(\Delta-\frac{d}{2}+1\right)}\,.
\ee 
}
\be
\frac{S_{m_j}^{\D_j}}{p_j^2 +\Delta_j+2m_j} 
%\frac{\Gamma\left(\Delta_j-\frac{d}{2}+1+m_j\right)}{2(m_j!) \Gamma^2\left(\Delta_j-\frac{d}{2}+1\right)}
\ee
where $\Delta_j$ is the dimension of the propagating scalar field.
\item Every vertex, $g \phi_1\dots \phi_k$ joining $k$ lines, contributes a factor\footnote{
The vertex factor is given by 
\be
V^{\Delta_1\dots \D_k}_{m_1 \dots m_k} =\sum_{n_1=0}^{m_1} \dots \sum_{n_k=0}^{m_k}
\Gamma\left(\frac{\sum_{j} (\Delta_j+2n_j) -d}{2}\right)
\prod_{j=1}^k \frac{(-m_j)_{n_j}}{n_j! \left(\D_j-\frac{d}{2}+1\right)_{n_j}}\,.
\nonumber
\ee}
\be
g\,V^{\Delta_1\dots \D_k}_{m_1 \dots m_k}
\ee 
\item Sum over all integers $m_j$ associated with internal lines. Each sum runs from 0 to $\infty$.
\item Multiply by 
\be
\mathcal{N}=\frac{ \pi^{\frac{d}{2} }} {2}\prod_{i=1}^n \frac{\sqrt{\mathcal{C}_{\D_i}}}{\Gamma(\D_i)}
\label{eq:constantN}
\ee
to get the $n$-point Mellin amplitude in our normalization of the external operators.
\end{itemize}

\begin{figure}
\begin{centering}
\includegraphics[clip,width=0.5\textwidth]{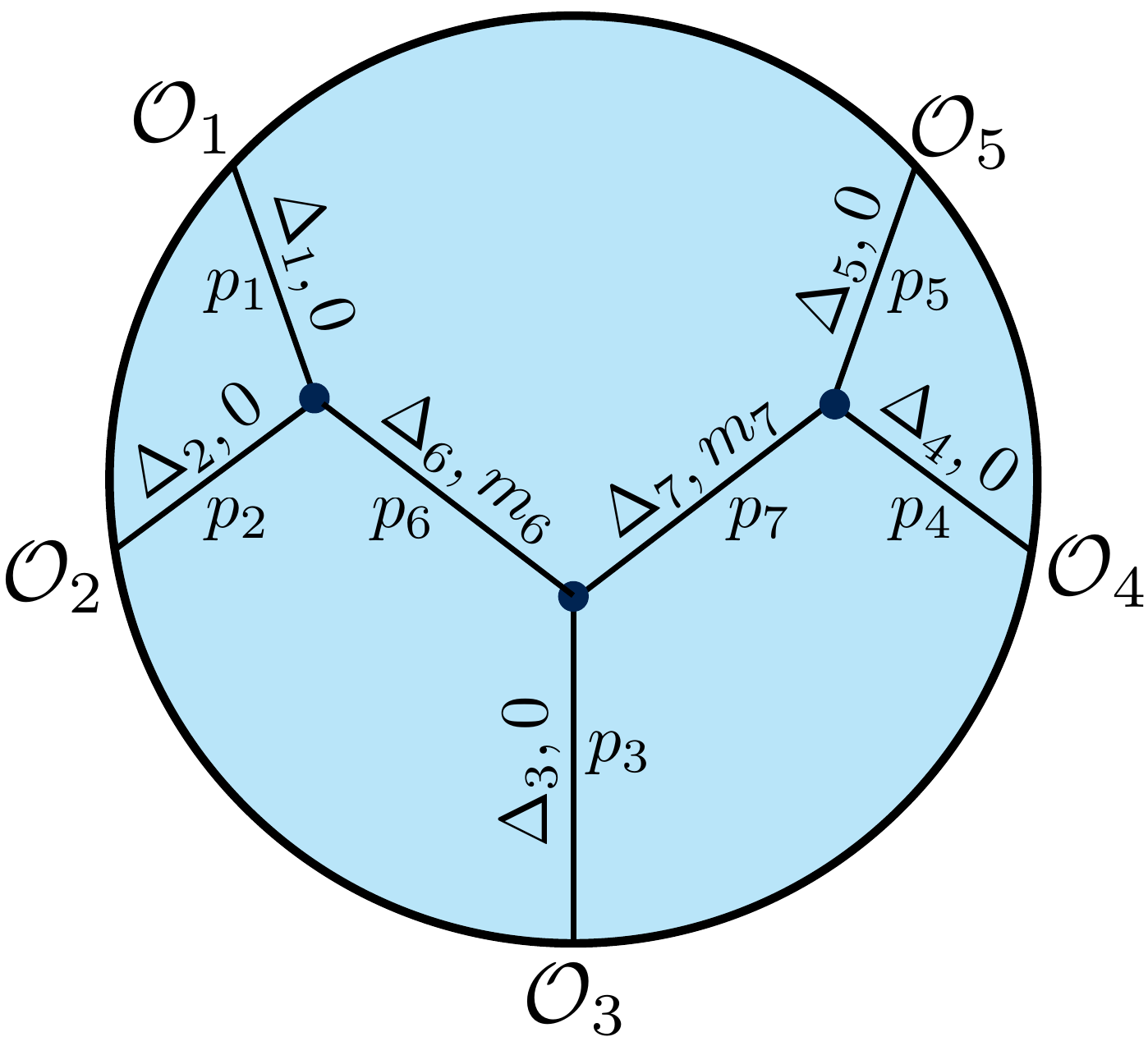}
\caption{A tree level scalar Witten diagram contributing to a 5-point function. 
The auxiliary momenta $p_i$ is conserved at each vertex, i.e. $p_6=p_1+p_2$ and $p_7=p_4+p_5$.
\label{fig:5pointdiagram}}
\end{centering}
\end{figure}

As an example, the Witten diagram in figure \ref{fig:5pointdiagram} gives rise to the following Mellin amplitude
\begin{align}
&\mathcal{N}\sum_{m_6=0}^\infty \sum_{m_7=0}^\infty 
 V^{\Delta_1\D_2 \D_6}_{0\,0\, m_6}
\frac{S_{m_6}^{\D_6}}{p_6^2 +\Delta_6+2m_6} 
V^{\Delta_6\D_3 \D_7}_{m_6\,0\, m_7}
\frac{S_{m_7}^{\D_7}}{p_7^2 +\Delta_7+2m_7} 
V^{\Delta_7\D_4 \D_5}_{m_7\,0\, 0}
\nonumber
\end{align}
where $p_6^2=(p_1+p_2)^2=2\g_{12}-\D_1-\D_2$ and $p_7^2=(p_4+p_5)^2=2\g_{45}-\D_4-\D_5$.
These Feynman rules suggest that we should think of the Mellin amplitude as an amputated amplitude because the bulk to boundary propagators do not contribute. In the case of scalar tree level diagrams (with non-derivative interaction vertices), the only dependence in the Mellin variables $\gamma_{ij}$ comes from the bulk-to-bulk propagators. % (after expressing the  $p_j^2$ in terms of the Mellin variables $\gamma_{ij}$). 
It is not known how to generalize these Feynman rules for loop diagrams or tree-level diagrams involving fields with spin. There are partial results in literature \cite{Paulos:2011ie, Goncalves:2014rfa} but nothing systematic.
Mellin amplitudes are also useful in the context of weakly coupled CFTs. The associated Feynman rules for tree level diagrams were given in \cite{Nizami:2016jgt}.

\begin{exercise}
Consider the residue of the Mellin amplitude at the first pole ($m=0$) associated to a bulk-to-bulk propagator.
Show that the Feynman rules above are compatible with the factorization property (\ref{eq:facMellinm0}) of this residue.
{\bf Extra:} check  the factorization formula (\ref{eq:facMellinm}) for the satellite poles with $m>0$.
\end{exercise}

\subsubsection{Flat space limit of AdS \label{sec:FSL}}

If we consider a scattering process where all length scales are much smaller than the AdS radius $R$ then the curvature effects should be negligible.
Consider a relativistic invariant theory in flat spacetime with a characteristic length scale $\ell_s$ (this scale could come from a mass or from a dimensionful coupling). Then, a scattering amplitude $\mathcal{T}_n$ of $n$ massless scalar particles in this theory will depend on $\ell_s$ and on the relativistic invariants $k_i\cdot k_j$, where $k_i$ are the momenta of the external particles.
On the other hand, this theory in AdS will give rise to Mellin amplitudes that depend on the dimensionless parameter $\theta=R/\ell_s$ and the Mellin variables $\g_{ij}$.
We claim that these two quantities are related by 
\be
\frac{\mathcal{T}_n(\ell_s,k_i)}{\ell_s^{n\frac{d-1}{2}-d-1}}= 
 \lim_{\theta\to\infty} \frac{1}{\mathcal{N}}
\int_{\Gamma} \frac{d\alpha}{2\pi i} \alpha^{\frac{d-\sum\D_i}{2} } e^\alpha
\frac{M_n\left(\theta,\gamma_{ij}=\frac{\theta^2}{2\alpha} \ell_s^2\,k_i\cdot k_j \right)}{
\theta^{n\frac{1-d}{2}+d+1}}
\label{eq:FSLformula}
\ee
where the contour $\Gamma$ runs parallel to the imaginary axis and passes to the right of the branch point at $\alpha=0$ and to the left of all poles of $M_n$.
The powers of $\ell_s$ where introduced to make both sides of the equation dimensionless and the constant $\mathcal{N}$ was given in \eqref{eq:constantN}.
The external particles are massless in flat space but in AdS they can have any scaling dimension $\Delta_i$ of order 1.
We expect this equation to hold when both sides of the equation are well defined. In case the flat space scattering amplitude $\mathcal{T}_n$ is IR divergent, we expect that the limit $\theta \to \infty$ of the Mellin amplitude will not be finite.\footnote{It might be useful to think of large $\theta$ as an IR regulator for the scattering amplitude.}

\begin{exercise}
Consider the vertex $\lambda \nabla\dots \nabla \phi_1 \dots \nabla\dots\nabla \phi_n$ discussed in exercise \ref{ex:generalcontact} in $d+1$ spacetime dimensions. Start by writing the coupling constant $\lambda$ as a power $(\ell_s)^q$ of a characteristic length scale $\ell_s$ and determine the value of $q$.
Then, use the Mellin amplitude \eqref{Mellingeneralcontact} in the flat space limit formula \eqref{eq:FSLformula} and obtain the expected $n$-particle scattering amplitude  
\be
\mathcal{T}_n=\lambda \prod_{i<j} (-k_i\cdot k_j)^{\alpha_{ij}}\ .
\ee
\end{exercise}

The last exercise can be seen as a derivation of the flat space limit formula \eqref{eq:FSLformula}. The point is that a generic Feynman diagram can be written as a (infinite) sum of contact diagrams with any number of derivatives. This corresponds to integrating out the internal particles and replacing there effect by contact vertices among the external particles. Since formula \eqref{eq:FSLformula} works for any contact diagram it should work in general.
This has been tested in several explicit examples, including 1-loop diagrams \cite{arXiv:1011.1485, arXiv:1107.1499, arXiv:1112.4845}. In addition, the same formula was derived in \cite{arXiv:1111.6972} using a wave-packet construction where the scattering region was limited to a small flat region of AdS.

In principle, formula \eqref{eq:FSLformula} provides a non-perturbative definition of string theory scattering amplitudes in terms of SYM correlation functions.
However, we do not know how to directly compute SYM correlators at strong coupling. In practice, what we can do is to use formula \eqref{eq:FSLformula} in the opposite direction, \emph{i.e.} we can use known string scattering amplitudes in flat space to obtain information about the strong coupling expansion of SYM correlators \cite{arXiv:1011.1485, arXiv:1411.1675}.
If the external  particles are massive in flat space then formula  \eqref{eq:FSLformula} is not adequate. This case was studied in \cite{Paulos:2016fap}.

\subsection{Open questions}

The study of Mellin amplitudes is still very incomplete.
Firstly, it is important to understand in what conditions do we have a well defined analytic Mellin amplitude. For example, in free CFTs the Mellin representation requires some form of regularization. This might be a  technical detail but it would be useful to understand in general the status of the Mellin representation.
Another important question is the asymptotic behavior of the Mellin amplitude when the Mellin variables are large. In the case of the four-point Mellin amplitude discussed in \ref{sec:MellinOPE4pt}, the limit of large $s$ with fixed $t$ is called the Regge limit in analogy with  flat space scattering amplitudes.
In \cite{arXiv:1209.4355}, we studied this limit using Regge theory techniques and making some reasonable assumptions about the large spin behaviour of the conformal partial amplitudes. Proving these assumptions is an important open question.
The bound on chaos \cite{Maldacena:2015waa} is another possible approach to the Regge limit of Mellin amplitudes.
Notice that if we can tame the asymptotic behaviour of $M(s,t)$ when $s\to \infty$, then we can write a dispersion relation that expresses $M(s,t)$ in terms of its poles in $s$, which are given by \eqref{eq:polesMellin4pt}.
This could provide a reformulation of the conformal bootstrap approach.

In the holographic context, it would be interesting to establish more general Feynman rules for Mellin amplitudes associated to Witten diagrams involving loops and particles with spin. 
It would also be useful to generalize more modern approaches to scattering amplitudes, like BCFW \cite{hep-th/0501052} or CHY \cite{arXiv:1307.2199}, to Mellin amplitudes.

\section*{Acknowledgements}
I would like to thank the students of TASI 2015 for their questions and enthusiasm. 
I am also grateful to the organizers Joe Polchinski, Pedro Vieira, Tom DeGrand and Oliver DeWolfe for their invitation and their patience with my delays in finishing these notes. 
\emph{Centro de F\'{i}sica do Porto} is partially funded by FCT. The research leading to these results has received funding from the People Programme (Marie Curie Actions) of the European Union's Seventh Framework Programme FP7/2007-2013/ under REA Grant Agreement No 317089 and CERN/FIS-NUC/0045/2015.

\ifarxivsubmission
\bibliographystyle{utphys}
\else
\bibliographystyle{ws-rv-van}
\fi

\bibliography{AdSCFTintro}

\ifarxivsubmission
\else
  \blankpage
\fi
\printindex
\end{document}